\def\sqr#1#2{{\vcenter{\vbox{\hrule height.#2pt\hbox{\vrule width.#2pt 
height#1pt \kern#1pt \vrule width.#2pt}\hrule height.#2pt}}}}
\def\d{\partial}

\def\w{\mathchoice\sqr45\sqr45\sqr{2.1}3\sqr{1.5}3\,} 
\def\=d{\,{\buildrel\rm def\over =}\,}

\documentclass[12pt]{article}
\usepackage{amssymb}\usepackage{amsmath}\usepackage{amsthm}
\newcommand{\CC}{\mathbb C}
\newcommand{\RR}{\mathbb R}

\newcommand{\NN}{\mathbb N}
\newcommand{\MM}{\mathbb M}
\newcommand{\supp}{{\rm supp\>}}

\newtheorem{prop}{Proposition}

\newtheorem{definition}{Definition}
\begin{document}
\title{The Master Ward Identity and \\ 
Generalized Schwinger-Dyson Equation in \\
Classical Field Theory}
\author{Michael D\"utsch
\thanks{Work supported by the Deutsche 
Forschungsgemeinschaft.} \\[2mm] 
Institut f\"ur Theoretische Physik\\
D-37073 G\"ottingen, Germany\\
{\tt \small duetsch@theorie.physik.uni-goettingen.de}\\[2mm] 
\and Klaus Fredenhagen \\[2mm]
II. Institut f\"ur Theoretische Physik\\
D-22761 Hamburg, Germany\\
{\tt \small klaus.fredenhagen@desy.de}}
\date{}
\maketitle
\begin{abstract}
In the framework of perturbative quantum field theory   
a new, universal renormalization condition (called Master Ward 
Identity) was recently proposed by one of us (M.D.) in a joint paper 
with F.-M. Boas. The main aim of the present 
paper is to get a better understanding of the Master Ward 
Identity by analyzing its meaning in classical field theory. 
It turns out that it is the most general identity for 
classical local fields which follows from the field equations. 
It is equivalent to a generalization of the Schwinger-Dyson Equation
and is closely related to the Quantum Action Principle of
Lowenstein and Lam. 

As a byproduct we give a self-contained treatment of  
Peierls' manifestly covariant definition of the Poisson 
bracket.

\noindent{\bf PACS.}{\small 11.10.Cd Field theory: axiomatic approach, 
11.10.Ef Field theory:
Lagrangian and Hamiltonian approach,
11.10.Gh Field theory: Renormalization,
11.15.Bt Gauge field theories: General properties of perturbation theory,
11.15.Kc Gauge field theories: Classical and semiclassical techniques}
\end{abstract}
\tableofcontents
\section{Introduction}
The hard question in the renormalization of a perturbative 
quantum field theory (QFT) is 
whether the symmetries of the underlying classical theory
can be maintained in the process of renormalization. The difficulties
are connected with the singular character of quantized fields which
forbids a straightforward transfer of the arguments valid for the
classical theory. Typically the various symmetries which one wants 
to be present in the
quantized theories are implied by certain identities (the Ward
identities) which one imposes as renormalization conditions. 

Traditionally, the impact of symmetries of the classical 
theory on the structure of quantum theory was analyzed in 
terms of the functional formulation of QFT (see 
\cite{Lam1,Lam2,Lo1,Lo2} and, e.g. \cite{Piguet}). 
In order to avoid infrared problems, it is, however, preferable to 
focus, in the spirit of algebraic quantum field theory, on the 
algebra of interacting fields. Actually, this becomes mandatory 
for quantum field theories on generic curved spacetimes (see, e.g. 
\cite{BF} and \cite{HW}). In the functional approach,
the algebraic properties of the interacting 
fields are not immediately visible. In causal perturbation 
theory \`{a} la Bogoliubov-Epstein-Glaser \cite{BS,EG}, on the other hand, 
the local algebras of observables of the interacting theory can 
be constructed directly \cite{BF} and, hence, we work with this method. 

In causal perturbation theory a general treatment of symmetries 
is the Quantum Noether Condition (QNC) of Hurth and Skenderis 
\cite{Hurth}. It addresses the problem: given a free classical 
theory with a symmetry, find a deformed 
classical Lagrangean which possesses a deformed symmetry, and extend 
the symmetry to the quantized theory.
In case of the BRS-current the QNC is closely related to 
the 'perturbative gauge invariance' of \cite{DHKS},
see the last Remark in Sect.~4.5.2 of \cite{DF3} (published version).

The main motivation for our works \cite{DF}, \cite{DF3} and to a 
certain extent for this paper is to give a construction of the local 
algebras of observables in quantum gauge theories, i.e. the elimination
of the unphysical fields and the construction of physical states in 
the presence of an adiabatically switched off interaction. In \cite{DF}
this construction was performed under certain conditions 
which were shown to be satisfied in QED. However, 
it turned out that in non-Abelian gauge theories 
additional relations, beyond QNC and perturbative gauge invariance, had to be 
fulfilled, in order to allow a local construction of the algebra of 
observables.

The Master Ward Identity (MWI) (postulated in \cite{DF3}) is a
universal formulation of symmetries. In \cite{DF3} it was 
shown that the MWI implies field equations, energy momentum 
conservation, charge conservation and a rigorous substitute for 
equal-time commutation relations of quark currents. Application of the 
MWI to the ghost- and to the BRS-current 
of non-Abelian gauge theories yields ghost number conservation
and the 'Master BRST Identity' \cite{DF3}. These symmetries contain
the information which is needed for our local construction of 
non-Abelian gauge theories.

Reference \cite{DF3} addresses the following problem.
\begin{itemize}
\item [(I)] Given a free theory, find a mapping from free fields to 
interacting fields, as a function of an interaction.
\end{itemize}
The MWI was obtained there in the following way: the difference
between different orders of differentiation and time-ordering,
\begin{equation}
    \d^\nu_{x_1} \tilde T(W_1,...,W_n)(x_1,...,x_n)- 
    \tilde T(\d^\nu W_1,...,W_n)(x_1,...,x_n)\label{[d,T]}
\end{equation}
($\tilde T(W_1,...,W_n)(x_1,...,x_n)$ denotes the time-ordered product 
of the Wick polynomials $W_1(x_1),...,W_n(x_n)$ in free fields), 
is formally computed by means of the
Feynman rules and the causal Wick expansion (see Sect.~4 of \cite{EG})
(or equivalently the normalization condition {\bf (N3)} \cite{DF}). 
The MWI requires then that renormalization has to be done in
such a way that this heuristically derived result is  
preserved. The main motivations for imposing this condition 
were, on the one 
hand side, the many, important
and far-reaching consequences of the MWI, and on the other 
hand side, the experience that the
MWI can nearly always be fulfilled. 

In this paper we give a further
important argument in favor of the MWI: {\it it is the straightforward
generalization to QFT of the most general classical identity 
for local fields which can be
obtained from the field equations and the fact that classical fields
may be multiplied point-wise} (see (\ref{fact1})). 
Since quantum fields are distributions which cannot, in 
general, be multiplied point-wise, the derivation of the MWI 
in classical field theory is not transferable to quantum 
field theory. There, the MWI is a highly
non-trivial normalization condition which contains much more
information than merely the field equations.
In order that the MWI is a well-defined and solvable renormalization 
condition also in models with anomalies it must be generalized, a
certain deviation from classical field theory must be admitted.
This has been worked out in Sect.~5 of \cite{DF3}.

The present paper is mainly dedicated to the problem:
\begin{itemize}
\item [(II)]  given a classical theory, find an appropriate dictionnary, 
relating classical fields to quantum fields.
\end{itemize}
We formulate 
identities in classical field theory in such a way that they remain 
meaningful in quantum field theory; concerning symmetries these 
identities are classical versions of the Schwinger-Dyson equations.  
As far as we know, our formulation is new (see however the 
papers \cite{deW,Mar}).

As just indicated
we start our study of the MWI with another equation, which will
turn out to be equivalent to the MWI. Namely we first formulate the
most general identity which follows in classical field theory from the
field equations. Due to formal similarity we call it the
{\it Generalized Schwinger-Dyson Equation} (GSDE). 
In this form it does not depend on a splitting of the 
Lagrangian into a free and an interaction part. We then 
introduce such a splitting and obtain the perturbative 
version of the GSDE which can be imposed as renormalization condition. 

It turns out that it is appropriate to use an off shell 
formalism where the entries of time-ordered products are 
classical fields not subject to any field equation, as 
advocated by Stora \cite{AWI}. 
{\it The MWI then gives a formula for 
time-ordered products of fields where one of the entries 
vanishes if the  
free field equations are imposed.}  
In the traditional version of causal perturbation theory 
all calculations are done in 
terms of Wick products of free fields. There the same 
identity becomes visible as a non-commutativity of 
differentiation and time-ordering (\ref{[d,T]}). 
In order to understand the 
connection between both formalisms we introduce a map 
$\sigma$ which associates free fields to general (off 
shell) fields. The time-ordered products $T$ of off shell 
fields in this paper are then related to the 
time-ordered products $\tilde{T}$ (\ref{[d,T]}) of on shell fields by 
\begin{equation}
        \tilde{T}(W_{1}(x_{1}),\ldots,W_{n}(x_{n}))=
        T(\sigma(W_{1})(x_{1}),\ldots,\sigma(W_{n})(x_{n}))\ .
\label{T:on/off-shell}
\end{equation}
In contrast to $\tilde{T}$, there 
is no reason why $T$ should not commute with derivatives. 
Therefore, we adopt here the proposal of Stora \cite{AWI} 
and postulate 
that $T$ can be freely commuted with derivatives. Stora calls 
this the Action Ward Identity (AWI), because it means that the 
interacting fields as well as the S-matrix depend on the 
interaction Lagrangian only via its contribution to the 
action.\footnote{It might be that 
Lemma 1 in \cite{Lo2} or Lemma 1 in \cite{Lam1} actually implies 
the AWI, but due to the rather different formalisms this conjecture 
could not yet be verified.} 

With that the non-commutativity of $\tilde T$ with derivatives 
is traced back to the non-commutativity of $\sigma$ with derivatives.
So, the MWI of \cite{DF3} can be formulated in terms of time ordered 
products where one of the entries is of the form 
$[\d_{\mu},\sigma](W)$. The latter expression vanishes if the 
free field equations are imposed, hence the MWI of \cite{DF3} 
is a special case of the MWI proposed in this paper. 
Actually, under a natural condition on the choice of 
$\sigma$, the two formulations are even equivalent. 

A puzzling observation was that the MWI of \cite{DF3} seemed to 
provide renormalization conditions already on the tree level, 
involving, in general, free parameters, in spite 
of the fact that the classical theory is unique. The solution
to this puzzle is that the map $\sigma$ is non-unique:
the freedom in the choice of parameters in the Feynman propagators of
derivated fields (see \cite{DHKS} and \cite{DF3}) 
is converted in the present formalism into the freedom in the choice of 
$\sigma$. Formula (\ref{T:on/off-shell}) can be interpreted in the
following way:
a fixed choice of $\sigma$ gives a solution of problem (I)
in terms of a solution of problem (II). 

The use of off shell fields has another advantage: it 
facilitates the introduction of auxiliary fields which 
in the presence of derivative couplings or in the 
definition of the BRS transformation may lead 
to a more elegant formulation. On the other hand, the use of 
auxiliary fields introduces more free parameters in the choice of $\sigma$.
 
Our analysis might be compared with the formulation of the Quantum Action
Principle of Lowenstein \cite{Lo1,Lo2} and Lam \cite{Lam1,Lam2}. These 
authors showed in the framework of
BPHZ renormalization how classical symmetries can be transferred into
renormalized perturbation theory. In contrast to these works, we emphasize
the structural similarity of classical and quantum perturbative field
theory. As a consequence, our arguments do not rely on the rather involved
combinatorics of BPHZ renormalization. However, we did not yet investigate
the structure of anomalies of the MWI. Another difference is that the
formalism of Lam seems to be inconsistent for vertices containing
higher than first derivatives of the basic fields, see Sect.~V
of \cite{Lam1}. We overcome these difficulties by means of the 
map $\sigma$ (cf. the Example (\ref{ex-MWI'})). To the best 
of our knowledge, a general prescription for the 
renormalization of terms with higher order derivatives does not seem 
to exist in the framework of causal perturbation theory 
(prior to \cite{DF3}, cf. the remark by Stora in \cite{AWI}).\footnote{The
treatment of vertices containing higher than first derivatives is not
an academic problem, e.g. the free BRS-current contains second
derivatives of the gauge fields.}

The paper is organized as follows: in Sect.~2.1 we study the 
canonical structure of classical field theory. 
We use Peierls' covariant definition of the
Poisson bracket \cite{P} which does not rely on a Hamiltonian
formalism.
In Sect.~2.2 we determine the perturbative expansion of the classical
fields as formal power series. The coefficients of this 
expansion are the retarded products. We prove that they 
satisfy the GLZ relations \cite{GLZ}. We briefly discuss the 
possibility of eliminating derivative couplings by 
introducing auxiliary fields in Sect.~2.3.

In Sect.~3 we formulate the GSDE and the MWI, introduce the map 
$\sigma$ and discuss the relation to the formulation of the 
MWI in \cite{DF3}.

In Sect.~4 we introduce the perturbative expansion of the
interacting quantum fields (as formal power series) by the principle
that {\it as much as possible of the classical structure is maintained in
the process of quantization}, in particular the GLZ 
relation, the AWI and the MWI.
All these conditions are formulated in terms of the retarded 
products, because in this formulation the conditions have 
the same form in the classical theory as well as in the 
quantum theory. In quantum theory, on the other hand, one 
can equivalently formulate everything in terms of the more 
familiar time-ordered products. 
The resulting formalism is not completely
equivalent to the one given in \cite{DF3}.
We clarify the significance of the difference.

In Sect.~5 we derive the
'Master BRST Identity' \cite{DF3} (which results from the application
of the MWI and AWI to the free BRS-current) in the formalism of 
this paper. We also use the MWI and AWI to determine the admissible
interaction of a BRS-invariant local gauge theory. By a modification 
of this procedure one can derive the conditions which are used in 
\cite{DF3} to express BRS-invariance of the interaction from more
fundamental principles. This is done in Appendix B. As a byproduct this
will clarify the relation to perturbative gauge invariance (in the 
sense of \cite{DHKS}).

Appendix A gives an explicit formula for the map 
$\sigma$ and shows its uniqueness in a particular framework.  
\section{Classical field theory for localized interactions}
\subsection{Retarded product and Poisson bracket}
To keep  the notations simple we consider only one real 
scalar field $\varphi$ (on the $d$-dimensional
Minkowski space $\MM$, $d>2$) and Lagrangians
\begin{equation}
\mathcal{L}=\mathcal{L}_0+\mathcal{L}_\mathrm{int}
\end{equation}
where the free part $\mathcal{L}_0$ is fixed. We will vary the 
interaction part 
\begin{equation}
\mathcal{L}_\mathrm{int}=-gP(\varphi,\d_\mu\varphi)
\=d :-gL_\mathrm{int},
\end{equation}
which is a polynomial $P=L_\mathrm{int}$ in $\varphi$ and 
$\d_\mu\varphi$ (later we will also allow for higher derivatives of
$\varphi$) multiplied by a test function $g\in\mathcal{D}(\MM)$. 
The latter is interpreted as a
space-time dependent coupling constant. We assume that the 
Cauchy problem is well posed for all Lagrangians in our class.
For simplicity we restrict our formalism to smooth solutions. 
In non-linear theories,
classical fields which are initially smooth may get singularities.
But, in this paper, we are mainly interested in perturbation theory.
It follows from the analysis in Sect.~2.2 that there is a unique 
smooth perturbative
solution, if the given incoming free solution is smooth.

Let $\mathcal{C_L}$ be the 
set of smooth solutions $f:\MM\rightarrow\RR$
of the Euler-Lagrange equation
\begin{equation}
        \partial_{\mu}\frac{\partial{\mathcal{L}}}
{\partial(\partial_{\mu}\varphi)} 
        =\frac{\partial{\mathcal{L}}}{\partial\varphi}\ ,\label{fieldeq}
\end{equation}
with compactly supported Cauchy data. We consider $\mathcal{C_L}$ as the 
classical phase space.
(This is equivalent to the traditional point of view in which an 
element of the phase space is the set of the corresponding Cauchy data, 
e.g. the functions $(f,\dot{f}),\>f\in \mathcal{C_L}$,  restricted 
to the time $x^0=0$.)

We interpret the field $\varphi$ as the evaluation functional on 
$\mathcal{C}\=d\mathcal{C}^\infty (\MM,\RR)$:
\footnote{A {\it complex} scalar field is the
analogous evaluation functional on $\mathcal{C}^\infty (\MM,\CC)$
and we define $\varphi^*(x)(f)\equiv f^*(x)$.}
\begin{equation}
\varphi(x)(f)\=d f(x),\quad\quad f\in \mathcal{C}^\infty 
(\MM,\RR) \ .
\label{eval:x}
\end{equation}
Functionals of the field
\begin{equation}
F(\varphi)\equiv\sum_{n=0}^N
\int dx_1...dx_n\,
\varphi (x_1)\ldots
\varphi (x_n)t_n(x_1,...,x_n),\quad N<\infty,
\label{F(fi)}
\end{equation}
then lead in a natural way to functionals on $\mathcal{C}$,
\begin{equation}
F(\varphi)(f)\=d F(f),\quad f\in \mathcal{C}.
\label{eval}
\end{equation}
Here $t_0\in\CC$ and the $t_n$ are suitable test functions
where we admit also certain distributions with compact support,
in particular $\delta^{4(n-1)}(x_1-x_n,...,x_{n-1}-x_n)f(x_n),\>f\in
{\cal D}(\MM)$. More precisely, we admit all distributions 
with compact support whose Fourier transform decays rapidly 
outside of the hyperplane $\{(k_{1},\ldots,k_{n}), \sum_i 
k_{i}=0\}$.  
We denote the algebra of functionals 
of the form (\ref{F(fi)}) by ${\mathcal F}({\mathcal C})$ and, when 
restricted to ${\mathcal C}_{\mathcal L}$, by   
${\mathcal F}({\mathcal C}_{\mathcal L})$. 

The field $\varphi_\mathcal{L}$ which satisfies the field 
equation (\ref{fieldeq}) is obtained
as the restriction of $\varphi$ to $\mathcal{C_L}$,
\begin{equation}
\varphi_\mathcal{L}\=d \varphi\vert_\mathcal{C_L},\label{fi_L}
\end{equation}
This restriction induces a homomorphism of the algebras of 
functionals $F$
\begin{equation}
F(\varphi)\to F(\varphi)_\mathcal{L}=F(\varphi_\mathcal{L}).\label{F}
\end{equation}
In particular, the factorization property
\begin{equation}
(AB)_\mathcal{L}(x)=A_\mathcal{L}(x)B_\mathcal{L}(x)\label{fact1}
\end{equation}
holds 
for polynomials $A,B$ in $\varphi$ and their partial derivatives. 
(The algebra of these polynomials will be denoted by ${\mathcal P}$,
\begin{equation}
  \mathcal{P}=\bigvee \{\d^a\varphi, a\in\NN_0^d\}\ . \ )\label{P}
\end{equation} 
It is a main difficulty of quantum field theory 
that the factorization property (\ref{fact1}) is no longer 
valid.

To compare theories with different Lagrangians we use the 
fact that by the assumed uniqueness of the solution of the 
Cauchy problem
there exists to each $f_2\in \mathcal{C}_{\mathcal{L}_2}$ precisely one 
$f_1\in \mathcal{C}_{\mathcal{L}_1}$ which coincides with $f_2$ 
outside of the future of the region where the respective Lagrangians 
differ,  
$f_1(x)=f_2(x)\>\forall x\not\in
(\supp (\mathcal{L}_{1}-\mathcal{L}_{2})
+\overline{V}_{+})$.\footnote{$V_{\pm}$ denote as 
usual the forward and backward 
light-cones, respectively, and 
$\overline{V}_{\pm}$ 
  their closures.} 
We denote the corresponding map (the 'wave operator')
by $r_{\mathcal{L}_1,\mathcal{L}_2}$:
\begin{equation}
   r_{\mathcal{L}_1,\mathcal{L}_2}:\mathcal{C}_{\mathcal{L}_2}\rightarrow
   \mathcal{C}_{\mathcal{L}_1}\ ,\ f_2\mapsto f_1.
   \label{r}
\end{equation}
Obviously it holds $r_{\mathcal{L}_1,\mathcal{L}_2}\circ
r_{\mathcal{L}_2,\mathcal{L}_3}=r_{\mathcal{L}_1,\mathcal{L}_3}$.
Analogously we define $a_{\mathcal{L}_1,\mathcal{L}_2}:
\mathcal{C}_{\mathcal{L}_2}\rightarrow\mathcal{C}_{\mathcal{L}_1},
f_2\mapsto f_1$ by requiring that $f_1$ and $f_2$ agree
in the distant future.

This bijection between the spaces of solutions can be used 
to express the interacting fields as functionals on the space of 
{\it free} solutions. We call
\begin{equation}
   A^\mathrm{ret}_{\mathcal{L}_\mathrm{int}}(x)\=d
   A(x)\circ r_{\mathcal{L}_0+\mathcal{L}_\mathrm{int},\mathcal{L}_0}:
   \mathcal{C}_{\mathcal{L}_0}\rightarrow \RR
   \label{fi^ret}
\end{equation}
for $A \in\cal{P}$ the 'retarded field'. The retarded
field is a functional on the free solutions which solves the 
interacting field equation. We will define
the perturbation expansion of classical interacting fields as the Taylor 
series of the retarded fields as functionals of the interaction Lagrangian,
\begin{equation}
  \label{eq:pert class field}
  A^\mathrm{ret}_{\mathcal{L}_{\mathrm{int}}}(x)= 
  \sum_{n=0}^{\infty} \frac{1}{n!}
  R_{n,1}(\mathcal{L}_{\mathrm{int}}^{\otimes n},A(x))\ .
\end{equation}
The retarded products $R_{n,1}$ will be constructed in
Sect.~2.2. Note
\begin{equation}
   \d^\mu_x \,A^\mathrm{ret}_{\mathcal{L}_\mathrm{int}}(x)=
   (\d^\mu A)^\mathrm{ret}_{\mathcal{L}_\mathrm{int}}(x)
   \label{dA^ret}
\end{equation}
and that the factorization property (\ref{fact1}) holds for the retarded
fields, too.

Besides the commutative and associative product (\ref{fact1}) there
is a second product for classical fields: the Poisson bracket.
Peierls \cite{P} has given a definition of 
the Poisson bracket 
without recourse to a
Hamiltonian formalism. We now review his procedure. It is 
convenient to 
generalize our formalism somewhat: we admit also non-local interactions,
i.e. the interaction part of the action $S$ does not need to be of 
the form $S_{\mathrm{int}}=\int
dx\,\mathcal{L}_{\mathrm{int}}(x)$,  
but may be replaced by an 
arbitrary functional $F\in\mathcal{F}(\mathcal{C})$.
The field equations are still obtained by the principle of least
action, but in contrast to (\ref{fieldeq}) they may involve non-local
terms. The classical phase space $\mathcal{C_L}$, 
the field $\varphi_\mathcal{L},
\,F(\varphi)_\mathcal{L}$  and the maps 
$r_{\mathcal{L}_1,\mathcal{L}_2},\,a_{\mathcal{L}_1,\mathcal{L}_2}$ 
(\ref{r}) are defined in the same
way as before, but now denoted by $\mathcal{C}_S,\,\varphi_S,
\,F(\varphi)_S$ and $r_{S_1,S_2},\,a_{S_1,S_2}$. 
(\ref{F}) and the factorization (\ref{fact1}) hold still 
true. We will not discuss whether solutions of the general 
Cauchy problem for these non-local actions exist. It is 
sufficient for our purpose that perturbative solutions 
always exist and are unique.  

Let $F\equiv F(\varphi)$ and $G\equiv G(\varphi)$ be functionals from 
$\mathcal{F}(\mathcal{C})$. We introduce the retarded product 
$R_S(F,G)$ and the advanced product $A_S(F,G)$,
\begin{gather}
  R_S(F,G)\=d \frac{d}{d\lambda}\vert_{\lambda =0}
  G\circ r_{S+\lambda F,S} \ ,
  \label{R()}\\
  A_S(F,G)\=d \frac{d}{d\lambda}\vert_{\lambda =0}
  G\circ a_{S+\lambda F,S}
  \label{A()}
\end{gather}
which are functionals on $\mathcal{C}_S$. 
Note that the entries of the retarded and advanced products 
are unrestricted functionals of the field. In general 
it is not possible to replace them by their restriction to 
the space of solutions, as the following example shows:
let $S=\int dx\,\mathcal{L}_0(x)$ with $\mathcal{L}_0=\frac{1}{2}
\d_\mu\varphi\d^\mu\varphi$, $F=\int dx\,g(x)\square\varphi(x)$
and $G=\varphi(y)$. Then $\varphi(y)\circ r_{S+\lambda F,S}(f)=f(y)
+\lambda g(y)$ and hence $R_S(F,\varphi(y))=g(y)$, but $F_S=0$.

The retarded and advanced products have the following important 
properties \cite{deW,Mar}: 
\begin{prop}\label{prop:ret}
{\rm (a)} The retarded product can be 
expressed in terms of the retarded Green function,
\begin{equation}
        R_{S}(F,G)=-\int dx\, dy\, \Bigl(\frac{\delta G}{\delta \varphi(x)} 
        \Delta_S^{\mathrm ret}(x,y)
        \frac{\delta F}{\delta \varphi(y)}\Bigr)_S\ .
        \label{retarded}
\end{equation}
\noindent {\rm (b)} The advanced and retarded products are related by 
\begin{equation}
  A_S(F,G)=R_S(G,F)\label{A=R}\ .
\end{equation}
\end{prop}
\noindent Here 
$\Delta_S^{\mathrm ret}$ 
is the unique retarded Green function of 
$\frac{\delta^2  S}{\delta \varphi(x)\delta\varphi(y)}$ 
considered as an integral 
operator, i.e. 
it satisfies the equations
\begin{equation}
      \int dy\,\Delta_S^{\mathrm ret}(x,y)
      \frac{\delta^2 
        S}{\delta \varphi(y)\varphi(z)}=\delta
      (x-z)=\int dy\,\frac{\delta^2 
        S}{\delta \varphi(x)\varphi(y)}\Delta_S^{\mathrm ret}(y,z),
        \label{S'':inv}
\end{equation}
and (in case $S$ is local)
\begin{equation}
   \supp \Delta_S^{\mathrm ret}
   \subset\{(x,y)\>|\> x\in y+\overline{V}_{+}\}.
   \label{supp-S'':inv}
\end{equation}
For non-local interactions with compact support we may construct the retarded 
Green function (and also the retarded product) in the sense of formal 
power series.
 
Since $\frac{\delta^2  S}{\delta \varphi(x)\varphi(y)}$ is symmetric, it 
follows that
\begin{equation}
   \Delta_S^{\mathrm adv}(x,y)\=d 
   \Delta_S^{\mathrm ret}(y,x)\label{Delta:av-ret}
\end{equation}
is the advanced Green function. Similarly to the proof of (a) one
finds that the advanced product $A_S(F,G)$ fulfills (\ref{retarded})
with $\Delta_S^{\mathrm ret}$ replaced by $\Delta_S^{\mathrm adv}$.
This and (\ref{Delta:av-ret}) immediately imply (b). 
\bigskip

\noindent {\it Example:} 
The abstract formalism may be illustrated by the example
of a real Klein Gordon field with a polynomial interaction, 
$S=\int dx\,\frac{1}{2}[\d^\mu
\varphi\d_\mu\varphi -m^2\varphi^2-gP(\varphi)]$. 
We obtain $\frac{\delta^2 
S}{\delta \varphi (x)\delta\varphi (y)}(f)=-
(\square+m^2+g(x)P''(f(x)))\delta(x-y)$, 
and 
\begin{equation}
   \Delta_S^{\mathrm ret}
   (x,y)(f)=-\Delta^{\mathrm{ret}}(x,y;gP''(f)),
   \label{S'':inv:KG}
\end{equation}
is the unique retarded Green function of the Klein-Gordon operator 
with a potential $gP''(f)$, where $f$ is the classical field configuration on 
which the functionals are evaluated. 
\bigskip

We will use the formula (\ref{retarded}) as definition of the 
retarded product outside of the space of solutions $f\in\mathcal{C}_S$ 
(and analogously for the advanced products), as it was done by
Marolf \cite{Mar}. 
\begin{proof}
It remains to prove (a).\\
Let $f\in\mathcal{C}_S$ and $r_{S+\lambda F,S}(f)=f+\lambda h+
\mathcal{O}(\lambda^2)$. Then
\begin{equation}
  0=\frac{d}{d\lambda}\vert_{\lambda=0}\frac{\delta(S+\lambda F)}
{\delta\varphi (y)}(f+\lambda h)=\frac{\delta F}
{\delta\varphi (y)}(f)+\int dz\,\frac{\delta^2 
        S}{\delta \varphi(y)\varphi(z)}(f)h(z)
\end{equation}
and (in the case of a local action $S$) $h(z)=0$ if 
$z\not\in\supp (F)+\bar V_+$.
Hence,
\begin{equation}
        R_S(F,\varphi(x))(f)=h(x)=-\int dy\, 
        \Delta_S^{\mathrm ret}(x,y)
        \frac{\delta F}{\delta \varphi(y)}(f),
\end{equation}
and by means of 
\begin{equation}
    R_{S}(F,G)(f)=\frac{d}{d\lambda}\vert_{\lambda=0}G(\varphi)(f+\lambda
    h)=\int dx\,\frac{\delta G}{\delta \varphi(x)}(f)h(x)
\end{equation}
we obtain the assertion (\ref{retarded}). (In the case of a non-local 
action the condition on the support of $h$ has to be 
appropriately modified.) 
\end{proof}
\begin{definition} The {\bf Peierls bracket} associated to an action $S$ 
  is a product on $\mathcal{F}({\mathcal C})$
  with values in $\mathcal{F}({\mathcal C}_S)$ defined by
  \begin{equation}
    \{F,G\}_S\=d R_S(F,G)-R_S(G,F)=R_S(F,G)-A_S(F,G).
    \label{Poisb}
  \end{equation}
\end{definition}
The Peierls bracket depends only on the restriction 
of the functionals to the space of solutions, 
\begin{equation}
  \{F,G\}_S=\{F',G\}_S \text{ if } F_S=F'_S \ . 
  \label{eq:Peierls-onshell}
\end{equation}
Namely, let $F$ be a functional which vanishes on the space 
of solutions. This is the case if $F$ is of the 
form\footnote{We tacitly 
assume here that the ideal ${\mathcal J}_S$ generated by the 
field equation (i.e. the set of functionals of the form (\ref{J_S}))
is identical with the set of 
functionals which vanish on the space of solutions. 
This seems to be true in relevant cases. Otherwise, we have to replace 
the restriction map $F\mapsto F_S$ by the quotient map with respect to 
${\mathcal J}_S$.}
\begin{equation}
    F=\int d x\, G(x)\frac{\delta S}{\delta \varphi(x)}\ .\label{J_S}
\end{equation}
Then the retarded product with a functional $H$ is
\begin{multline}
        R_{S}(F,H)=\\
        \int dx\,dy\,dz\,
        \Bigl(
        \frac{\delta G(x)}{\delta\varphi(y)} 
        \frac{\delta S}{\delta \varphi(x)} 
        + G(x) \frac{\delta^2 
        S}{\delta\varphi(x)\delta\varphi(y)}\Bigr)
        \Delta_S^{\mathrm ret}(y,z) \frac{\delta H}{\delta\varphi(z)} \ .
\label{a30}
\end{multline}
The first term vanishes on the space of solutions, 
therefore we obtain
\begin{equation}
        R_{S}(F,H)= \int dz\, G(z)\frac{\delta H}{\delta\varphi(z)}\ .
\label{a31}
\end{equation}
The same expression is obtained for the advanced product, 
thus the Peierls
bracket of $F$ and $H$ vanishes.
We may therefore define the Peierls bracket for functionals on the space 
of solutions by
\begin{displaymath}
  \{F_S,G_S\}=\{F,G\}_S \ .\label{Pb}
\end{displaymath}
It is easy to see that for the example of the Klein Gordon field 
with a polynomial interaction without derivatives the Peierls 
bracket coincides with the Poisson bracket obtained from the 
Hamiltonian formalism. The Peierls bracket, however is defined also 
for derivative couplings and even for non-local interactions 
where the Hamiltonian formalism has problems. Moreover, it is 
manifestly covariant and does not use a splitting of space-time 
into space and time. 
  
We now want to show that the  Peierls bracket fulfils in general the usual 
properties of a Poisson bracket. Antisymmetry, linearity and the 
Leibniz rule are obvious
\footnote{Linearity and the Leibniz rule hold already for the
  retarded and advanced products.},
but the Jacobi identity is non-trivial (actually it is not 
discussed in the paper of Peierls, however in \cite{deW,Mar}). 
We will see that the Jacobi identity follows from the fact that 
$r_{S_2,S_1}$ commutes with the Peierls bracket (hence it is a
canonical transformation).
\begin{prop}\label{prop:Jacobi}
  {\rm (a)} The retarded wave operator 
  $r_{S_{2},S_{1}}$ preserves the Peierls bracket 
  (\ref{Poisb}),
  \begin{equation}
     \{F\circ r_{S_2,S_1},G\circ r_{S_2,S_1}\}_{S_1}=
     \{F,G\}_{S_2}\circ r_{S_2,S_1}
     \label{r-Pb}
  \end{equation}
  and the same statement holds for $a_{S_2,S_1}$.

  \noindent {\rm (b)} The Peierls bracket (\ref{Poisb}) satisfies 
  the conditions which 
  are required for a Poisson bracket, in particular the Jacobi identity 
  \begin{equation}
     \{F_S,\{H_S,G_S\}\}+\{G_S,\{F_S,H_S\}\}+\{H_S,\{G_S,F_S\}\}=0 \ .
     \label{Jacobi}
  \end{equation}
\end{prop}
\begin{proof}
  It suffices to prove (\ref{r-Pb}) for an infinitesimal
  change of the interaction: setting $S_1=S$ and $S_2=S+\lambda H$,
  the infinitesimal version of (\ref{r-Pb}) reads
  \begin{multline}
     \{R_S(H,F),G_S\}+\{F_S,R_S(H,G)\}=\\
     R_S(H,\{F,G\})+\frac{d}{d\lambda}\vert_{\lambda=0}
     (R_{S+\lambda H}(F,G)-A_{S+\lambda H}(F,G))
     \label{inf}
  \end{multline}
  where we used in the last term the extended definition of the retarded and 
  advanced 
  products outside of the 
  space of solutions introduced after Proposition~\ref{prop:ret}.

  We now insert the formula for the retarded product of 
  Proposition~\ref{prop:ret} everywhere in (\ref{inf}).
  Applying $\frac{\delta}{\delta\varphi}$ to (\ref{S'':inv}) we find 
  \begin{equation}
    \frac{\delta}{\delta\varphi (z)}
    \Delta_S^{\mathrm adv}(x,y)=
    -\int dv\,dw\,\Delta_S^{\mathrm adv}(x,v)
    \frac{\delta^3 S}{\delta\varphi (v)\delta\varphi (w)\delta\varphi (z)}
    \Delta_S^{\mathrm adv}(w,y),
  \end{equation}
  and analogously
  \begin{equation}
     \frac{d}{d\lambda}\vert_{\lambda=0}
     \Delta_{S+\lambda H}^{\mathrm adv}(x,y)=
     -\int dv\,dw\,\Delta_S^{\mathrm adv}(x,v)
     \frac{\delta^2 H}{\delta\varphi (v)\delta\varphi (w)}
     \Delta_S^{\mathrm adv}(w,y).
  \end{equation}
  With that (\ref{inf}) can be verified by a straightforward calculation. 

  By an analogous calculation we prove that the advanced transformation 
  $a_{S_2,S_1}$ is a canonical transformation. The infinitesimal version is 
  (\ref{inf}) where in the first three terms $R_S$ is replaced
  by $A_S$ and where the last term, the term with the $\lambda$-derivative, 
  is unchanged. Hence, considering
  the difference of these two versions of (\ref{inf}), the latter term 
  drops out, and we obtain the Jacobi identity (\ref{Jacobi}). 
\end{proof}
\subsection{Higher order retarded products and perturbation theory}
In analogy to equation (\ref{R()}) we define the higher order retarded 
products by
\begin{equation}\label{R_n()}
   R_S(F^{\otimes n},G)=
   \frac{d^n}{d\lambda^n}\vert_{\lambda=0}G\circ r_{S+\lambda F,S}\ .  
\end{equation}
They have a unique extension to $(n+1)$-linear functionals 
on ${\mathcal F}({\mathcal C})$ which are symmetric in the 
first $n$ variables. With that the perturbative expansion of
$G\circ r_{S+\lambda F,S}$ in $\lambda$ reads
\begin{equation}
  G\circ r_{S+\lambda F,S}\simeq\sum_{n=0}^\infty\frac{\lambda^n}{n!}
  R_S(F^{\otimes n},G)\equiv:R_{S}(e_{\otimes}^{\lambda F},G)
\label{pert-exp}
\end{equation}
in the sense of formal power series. If $S$ is the free part of the
action, this is the perturbative expansion of the retarded fields 
(\ref{fi^ret}) in terms of free
fields.
    
In the first paper of \cite{DF1}, equation (71), we gave an explicit 
formula for $R$ in the case where 
all functionals are local and where
$F$ and $G$ do not contain derivatives. 
It took the form of a retarded multi-Poisson bracket. In case of derivative 
couplings this can no longer be true, as we already know from the 
discussion of the retarded product of two factors, cf. the counter 
example after equation (\ref{A()}). 
\bigskip

The {\it general} case where $S$, $F$ and $G$ might be non-local 
can be obtained from the formula
\begin{equation}
        \frac{d}{d\lambda}R_{S}(e_{\otimes}^{\lambda 
        F},G)=R_{S}(e_{\otimes}^{\lambda F},R_{S+\lambda F}(F,G))
        \label{eq:ret-recurs}
\end{equation}
which is the Taylor series expansion of 
\begin{equation}
        \frac{d}{d\lambda}G\circ r_{S+\lambda F,S} = R_{S+\lambda 
        F}(F,G)\circ r_{S+\lambda F,S} \ ,
        \label{eq:ret-recurs1}
\end{equation}
the latter identity following directly from the definition of the retarded 
products. On the r.h.s. of (\ref{eq:ret-recurs}) we understand 
$R_{S+\lambda F}(F,G)$ as an unrestricted functional, i.e.
$R_{S+\lambda F}(F,G)\in\mathcal{F(C)}$. 
Comparing the coefficients on both sides of (\ref{eq:ret-recurs}) yields a 
recursion relation:
\begin{equation}
  R_S(F^{\otimes (n+1)},G)=-\sum_{l=0}^n \binom{n}{l}
R_S\Bigl( F^{\otimes l},\int dx\, dy\, \frac{\delta F}{\delta \varphi(x)} 
        \Delta_{S+\lambda F}^{\mathrm{adv}\,(n-l)}(x,y)
        \frac{\delta G}{\delta \varphi(y)}\Bigr),
\end{equation}
where
\begin{gather}
  \Delta_{S+\lambda F}^{\mathrm{adv}\,(k)}(x,y)\=d
\frac{d^k}{d\lambda^k}\vert_{\lambda =0}
\Delta_{S+\lambda F}^{\mathrm{adv}}(x,y)\notag\\
=(-1)^k k!\int dv_1...dv_k\, dz_1...dz_k\, 
\Delta_{S}^\mathrm{adv}(x,v_1)\notag\\
\cdot\frac{\delta^2 F}{\delta\varphi (v_1)\delta\varphi (z_1)}
\Delta_S^{\mathrm{adv}}(z_1,v_2)...
\frac{\delta^2 F}{\delta\varphi (v_k)\delta\varphi (z_k)}
\Delta_S^{\mathrm{adv}}(z_k,y)\>.\notag
\end{gather}
This shows the existence of solutions in the sense of 
formal power series.

Peierls' formula for the Poisson bracket together with the 
fact that the retarded wave operators $r_{S,S_{0}}$ are 
canonical transformations lead to an interesting relation 
between the higher order retarded products.
 
Let $F,G$ and $S_{1}$ be functionals from 
${\mathcal F}({\mathcal C})$ and let $S=S_{0}+\lambda S_{1}$. Then we have
\begin{equation}
        \{F\circ r_{S,S_{0}},G\circ r_{S,S_{0}}\}_{S_{0}} = 
        (R_{S}(F,G)-R_{S}(G,F))\circ r_{S,S_{0}} \ .
        \label{eq:GLZ1}
\end{equation}  
According to the definition of the retarded products it holds
\begin{displaymath}
        R_{S}(F,G)\circ 
        r_{S,S_{0}}=\frac{d}{d\mu}\vert_{\mu =0} G\circ 
        r_{S+\mu F,S_{0}}
\end{displaymath}
where we used the composition property of the wave operators. 
If we now take the $n$-th derivative with respect to 
$\lambda$ on both sides of 
(\ref{eq:GLZ1}) we find
\begin{gather}
     \sum_{I\subset \{1,\ldots,n\}}
     \{R_{S_{0}}(\otimes_{i\in I}H_{i},F),
     R_{S_{0}}(\otimes_{j\not\in I} H_{j},G)\}_{S_{0}}  
     =\notag\\
     R_{S_{0}}(\otimes_{i=1}^n H_{i}\otimes F,G)- 
     R_{S_{0}}(\otimes_{i=1}^n H_{i}\otimes G,F)
     \label{eq:GLZ2}
\end{gather}
with $H_{i}\in {\mathcal F}({\mathcal C})$.
This relation is in quantum field theory known as the 
GLZ-Relation (see below). It plays an important role in 
renormalization.
\bigskip 

In case $S$ and $F$ are {\it local}, we can find an elegant 
expression for the retarded products. We introduce the following differential 
operator on the space of functionals ${\mathcal F}({\mathcal C})$
\begin{equation}\label{eq:R-Op}
  {\mathcal R}(x):= 
  -\int dy\, \Bigl(\frac{\delta F}{\delta \varphi(x)} 
  \Delta_S^{\mathrm ret}(y,x)
  \frac{\delta }{\delta \varphi(y)}\Bigr)\ .  
\end{equation}
Note that ${\mathcal R}(x)$ is smooth in $x$ since $\Delta_S^{\mathrm ret}$ 
maps smooth functions with compact support onto smooth functions. 
According to (\ref{retarded}) we have
\begin{equation}
  \label{eq:R1}
  R_S(F,G)=\int dx\, ({\mathcal R}(x)G)_S \ .
\end{equation}
The $n$-th order case looks quite similar:
\begin{prop}\label{prop:Rn}
  The $n$-th order retarded product is given by the formula
  \begin{equation}
    \label{eq:Rn}
    R_S(F^{\otimes n},G)=n!\int_{x_1^0\le \ldots \le x_n^0}dx_1\ldots dx_n 
    ({\mathcal R}(x_1)\cdots {\mathcal R}(x_n)G)_S
  \end{equation}
\end{prop}
\begin{proof}
   We first show that the power series defined by the 
   r.h.s. of formula (\ref{eq:Rn}) 
   \begin{equation}
     \label{eq:interacting}
     G\mapsto G(\lambda)=R^0_S(\exp_{\otimes}\lambda F,G)
   \end{equation}
   defines a homomorphism on the algebra of functionals 
   $G\in{\mathcal F}({\mathcal C})$. This means
   that for two functionals $G$ and $H$ we have the factorization
   \begin{equation}
     \label{eq:factorization}
     R^0_S(F^{\otimes n},GH)=
     \sum_{k=0}^n \binom{n}{k}
     R^0_S(F^{\otimes k},G)R^0_S(F^{\otimes n-k},H) \ .
   \end{equation}
   We use the fact that the operators ${\mathcal R}(x)$ are 
   functional derivatives of first order. Hence from Leibniz' rule we get
   \begin{equation}
     \label{eq:Leibniz}
     {\mathcal R}(x_1)\cdots {\mathcal R}(x_n)GH = 
     \sum_{I\subset \{1,\ldots,n\}}
     (\prod_{i\in I}{\mathcal R}(x_i)G)(\prod_{j\not\in I}{\mathcal 
     R}(x_j)H) \ .
   \end{equation}
    It remains to check the time ordering prescription. For any 
    $n$-tupel of times $t=(x^0_1,\ldots,x^0_n)$ we choose a permutation $\pi_t$ 
    with
    $x^0_{\pi_t(1)}\le\ldots\le x^0_{\pi_t(n)}$. 
    We obtain
   \begin{equation}
     \label{eq:timeordering}
     R^0_S(F^{\otimes n},G)=\int d^n x ({\mathcal R}(x_{\pi_t(1)})\ldots 
     {\mathcal R}(x_{\pi_t(n)})G)_S \ .   
   \end{equation}
   If we insert (\ref{eq:Leibniz}) into (\ref{eq:timeordering}) we see 
   that the permutation $\pi_t$ restricted to $I$ as well as to the 
   complement of $I$ yields the correct time ordering. The $n$-fold 
   integral factorizes, and the integrals over $(x_i,i\in I)$ and 
   $(x_j,j\not\in I)$ do not 
   depend on the choice of  $I$, but only on the cardinality of $I$. 
   This proves (\ref{eq:factorization}).
    
   We now show that the formal power series for the retarded 
   field (\ref{eq:interacting}) with action $S+\lambda F$ satisfies the field 
   equation 
   $\frac{\delta}{\delta\varphi(x)}(S+\lambda F)=0$ and thus 
   coincides with $R_{S}(\exp_{\otimes}\lambda F,G)$ because 
   of (\ref{eq:factorization}) and the uniqueness of the retarded solutions. 
   We have to show  $R^0_S(\exp_{\otimes}\lambda F,\frac{\delta (S+\lambda F)}
   {\delta\varphi (x)})=0$. So we insert 
   $G=\frac{\delta}{\delta\varphi(y)}S $ into 
   (\ref{eq:interacting}), use
   \begin{equation}
         {\mathcal R}(x)\frac{\delta}{\delta\varphi(y)}S 
         =-\frac{\delta F}{\delta\varphi(x)}
         \int dz\, \Delta_{S}^{\mathrm ret}(z,x)
         \frac{\delta^2 S}{\delta\varphi(x)\delta\varphi(z)}=
         -\frac{\delta F}{\delta\varphi(x)}\delta(x-y)\ .
         \label{eq:fieldequation}
   \end{equation}
   and obtain the wanted field equation where we exploit the 
   fact that ${\mathcal R}(y)\frac{\delta F}{\delta\varphi(x)}=0$ if $y$ is 
   not in the past 
   of $x$. 
\end{proof}

Since we are mainly interested in local functionals, we 
change our point of view somewhat. Let ${\mathcal P}$ denote 
as before the set of polynomials of $\varphi$ and 
its derivatives (\ref{P}). 
Each field $A\in{\mathcal P}$ defines a distribution with values 
in ${\mathcal F}(\mathcal{C})$,
 \begin{displaymath}
        A(f)=\int dx\, A(x)f(x)\ ,\ f\in{\mathcal D}({\MM})\ .
 \end{displaymath}
We fix a local action $S$ which later will be the free 
action. We now define the retarded products of fields as 
${\mathcal F}({\mathcal C}_{S})$ valued distributions in several 
variables by
 \begin{gather}
        R_{n,1}(A_1\otimes\cdots\otimes A_n,B)
        (f_1\otimes\cdots\otimes f_n,g)\equiv\notag\\
        \int d(x,y)\, R_{n,1}(A_{1}(x_{1}),\ldots,A_{n}(x_{n});B(y))
        f_{1}(x_{1})\cdots f_{n}(x_{n})g(y)\=d \notag\\
        R_{S}(A_{1}(f_{1})\otimes \cdots \otimes A_{n}(f_{n}),B(g))
        \label{eq:retarded products of local fields}
\end{gather}
The retarded products $R_{n,1}$ are multi-linear functionals on ${\mathcal P}$ 
with values in the space of  ${\mathcal F}({\mathcal 
C}_{S})$ valued distributions. We may equivalently consider 
them as distributions on the space of ${\mathcal 
P}^{\otimes n+1}$ valued 
test functions ${\mathcal D}(\MM^{n+1},{\mathcal 
P}^{\otimes n+1})=({\mathcal 
P}\otimes {\mathcal D}(\MM))^{\otimes n+1}$, i.e. we sometimes write
$R_{S}(A_{1}f_{1}\otimes \cdots \otimes A_{n}f_{n},Bg)$  

In Sect.~4 we will define perturbative quantum fields by the principle
that as much as possible of the structure of perturbative classical
fields is maintained in the process of quantization. For this purpose we are
going to work out main properties of the retarded products $R_{n,1}$
(\ref{eq:retarded products of local fields}).
\begin{itemize}
        \item  The {\bf causality} of the retarded fields, 
\begin{equation}
   B_{S+A(f)}(x)= 
   B_{S}(x),\quad\textrm{if}\quad\supp f
   \cap (x+\overline{V_-})=\emptyset
   \label{caus:intf}
\end{equation}
(where $f\in\mathcal{D}(\MM), A,B\in\mathcal{P}$)
translates into the support property:
\begin{equation}
   {\rm supp}\>R_{n,1}\subset\{(x_1,...,x_n,x)|
    x_l\in x+\overline{V_-},\forall l=1,...,n\}.
   \label{supp(R)}
\end{equation}
\item  
   A deep property of the retarded products $R_{n,1}$ is the 
   {\bf GLZ-Relation}. In (\ref{eq:GLZ2}) we formulated it for 
   general retarded products. For the retarded products of 
   local fields the GLZ-Relation reads
   \begin{gather}
      \sum_{I\subset \{1,...,n\}}\{R_{|I|,1}(\otimes_{i\in I}f_i;f),
      R_{|I^c|,1}(\otimes_{k\in I^c}f_k;g)\}=\notag\\
      R_{n+1,1}(f_1\otimes...\otimes f_n\otimes f;g)-
      R_{n+1,1}(f_1\otimes...\otimes f_n\otimes g;f)
      \label{Pb(R)}
   \end{gather}
   for $f_1,...,f_n,f,g\in\mathcal{P}\otimes \mathcal{D}(\MM)$. 

   Glaser, Lehmann and Zimmermann (GLZ) 
   \cite{GLZ} found this formula in the framework of non-perturbative
   QFT for the retarded products introduced by Lehmann, Symanzik 
   and Zimmermann \cite{LSZ2}. In causal perturbation theory \cite{EG}
   the GLZ-relation is a consequence of Bogoliubov's definition of
   interacting fields \cite{BS}, see Proposition 2 in \cite{DF}.
   The important point is that the retarded 
   products on the l.h.s. in (\ref{Pb(R)}) are of lower orders, 
   $|I|,|I^c|< n+1$. 
   
\item  
   From their definition (\ref{eq:retarded products of local 
   fields}) it is evident that the {\bf retarded products $R_{n,1}$ 
   commute with partial derivatives}
   \begin{equation}
      \d^\mu_{x_{l}} R_{n,1}(\ldots,A_l(x_l),\ldots)=
      R_{n,1}(\ldots,\d^\mu A_l,\ldots) \ , \  
      l=1,\ldots,n+1\ ,
      \label{[d,R]}
   \end{equation}
   where $A_1,\ldots,A_{n+1}\in\mathcal{P}$. 
   Note that the kernel of the linear map
   \begin{equation}
      \mathcal{P}\otimes \mathcal{D}(\MM)\longrightarrow
      {\mathcal F}({\mathcal C}):\> 
      A\otimes g\mapsto A(g)
      \label{P>func}
   \end{equation} 
   is precisely the linear span of
   $\{\d^\mu A\otimes g + A\otimes \d^{\mu}g\>|\> A\in\mathcal{P},
   g\in\mathcal{D}(\MM)\}$. 
   (\ref{[d,R]}) expresses the fact
   that the retarded products $R_{n,1}$ depend on 
   the functionals (i.e. the images
   of the map (\ref{P>func})) only. 
   This can be interpreted in physical terms:
   Lagrangians which give the same action yield the same physics.
   This is the motivation for Raymond Stora to require (\ref{[d,R]})
   for the retarded (or equivalently: time ordered) products of QFT,
   and he calls this the {\bf Action Ward Identity} (AWI) \cite{AWI},
   see Sect.~4.

\item  
   We now assume that $S$ is at most quadratic in 
   the fields. Then the second derivative is independent 
   of the fields, and therefore also the Green functions. We set
   \begin{equation}
        \Delta_{S}(x,y)\=d  
        -\Delta_{S}^{\mathrm ret}(x,y)+ 
        \Delta_{S}^{\mathrm adv}(x,y) \ .
   \end{equation}
   We look at the {\bf Poisson bracket of a retarded product with a free
    field}. Let $A,B\in {\mathcal P}$ 
    and $f,g,h\in {\mathcal D}(\MM)$. We are interested in 
    \begin{equation}
        \{R_{n,1}(A^{\otimes n},B)(f^{\otimes 
        n},g),\varphi(h)\}_{S} \ .
    \end{equation}      
    By definition of the retarded products of local fields 
    this is equal to
    \begin{equation}
        \{R_{S}(A(f)^{\otimes n},B(g)),\varphi(h)\}_{S}
    \end{equation}
    We apply Proposition~\ref{prop:ret} to compute the Poisson
    bracket. From Proposition~\ref{prop:Rn} it follows 
    that $R_{S}$ commutes with functional derivatives if $S$ 
    is a quadratic functional. Hence we obtain
    \begin{gather}
        \int dx\,\int dy\, \Bigl(R_{S}(n\frac{\delta A(f)}{\delta 
        \varphi(x)}\otimes A(f)^{\otimes n-1},B(g)) +\notag\\ 
        R_{S}(A(f)^{\otimes n},\frac{\delta 
        B(g)}{\delta\varphi(x)})\Bigr) \Delta_{S}(x,y)h(y)
        \label{eq:N3preparation}
    \end{gather}
    Using the formula
    \begin{equation}
        \int dx\, \frac{\delta A(f)}{\delta\varphi(x)}k(x)=
        \sum_{a}\int dx\, \frac{\d 
        A}{\d(\d^{a}\varphi)}(x)f(x)\d^{a}k(x)
        \label{eq:derivative}
    \end{equation}
    for the functional derivative of a local functional, we 
    finally arrive at the formula
    \begin{gather}
        \{R_{n,1}(f_{1},\ldots,f_{n+1}) ,
        \varphi(h)\}_{S} =\notag\\
        \sum_{k=1}^{n+1}\sum_{a}(R_{n,1}(f_{1}, 
        \ldots ,\frac{\d 
        f_{k}}{\d(\d^{a}\varphi)}\d^{a}\Delta_{S}h,\ldots, 
        f_{n+1}) 
       \label{eq:N3class}
    \end{gather}
    where $f_1,\ldots,f_{n+1}\in\mathcal{D}(\MM,\mathcal{P})$, 
    $h\in\mathcal{D}(\MM)$ and where $\Delta_{S}$ was considered 
    as an integral operator 
    acting on $h$.
    
    The requirement that this relation
    holds also in perturbative QFT plays an important role in the 
    inductive construction of perturbative quantum fields (see 
    Sect.~4).

    \item  {\bf Symmetries:} there are natural automorphic actions 
    $\alpha_L$ and $\beta_L$ of the Poincare group 
    ($L\in {\cal P}_+^\uparrow$) on 
    $({\cal P}\otimes {\cal D}(\MM))^{\otimes n+1}$
    and on $\mathcal{F}(\mathcal{C})$, respectively. 
         The retarded products are {\bf Poincare covariant}: $R_{n,1}\circ
    \alpha_L=\beta_L\circ R_{n,1}$, provided $S$ is 
    invariant. 
    A universal formulation
    of all symmetries which can be derived from the field equations in
    classical field theory is given by the {\bf MWI}, see Sect.~3.
     
    \item 
    The {\bf factorization} $(AB)^{\mathrm{ret}}_{\mathcal{L}_\mathrm{int}}(x)
    =A^{\mathrm{ret}}_{\mathcal{L}_\mathrm{int}}(x)
     B^{\mathrm{ret}}_{\mathcal{L}_\mathrm{int}}(x)$
    and the Leibniz rule for the retarded product of two 
    factors
    yield, in general, ill-defined expressions in QFT. It is the 
    Master Ward Identity which allows to implement the 
    consequences of the factorization property of classical 
    field theory into quantum field theory.

\end{itemize}
\subsection{Elimination of derivative couplings} 
Interaction Lagrangians containing derivatives of fields usually cause
complications in the canonical formalism. They also change relations
between different fields, as may be seen by the non-linear term in the
formula which expresses the field strength $F^{\mu\nu}$ of a
Yang-Mills theory in terms of the vector potential $A^\mu$.
A convenient way to deal with these complications is the introduction
of auxiliary fields.

As an example we consider the Lagrangian
\begin{equation}
    \mathcal{L}(\varphi,\d^\mu\varphi)=\frac{1}{2}\d^\mu\varphi\d_\mu\varphi
    -\frac{m^2}{2}\varphi^2+\mathcal{L}_\mathrm{int}(\varphi,\d^\mu\varphi)
    \label{L1}
\end{equation}
of a real scalar field $\varphi$ with the Euler-Lagrange equation
\begin{equation}
    (\square +m^2)\varphi =\frac{\d\mathcal{L}_\mathrm{int}
    (\varphi,\d^\mu\varphi)}{\d\varphi}-\d^\nu\frac{\d\mathcal{L}_\mathrm{int}
    (\varphi,\d^\mu\varphi)}{\d\,\d^\nu\varphi}.\label{f3}
\end{equation}
To eliminate $\d^\mu\varphi$ in $\mathcal{L}_\mathrm{int}$
we introduce a vector field $\varphi^\mu$ and a Lagrangian 
\begin{equation}
    \mathcal{L}(\varphi,\d^\mu\varphi,\varphi^\mu)=-\frac{1}{2}
    \varphi^\mu\varphi_\mu +\varphi_\mu\d^\mu\varphi
    -\frac{m^2}{2}\varphi^2 +\mathcal{L}_\mathrm{int}(\varphi,\varphi^\mu).
    \label{L2}
\end{equation}
with the Euler-Lagrange equations:
\begin{gather}
     0=-\varphi_\mu +\d_\mu\varphi +\frac{\d\mathcal{L}_\mathrm{int}
     (\varphi,\varphi^\mu)}{\d\varphi^\mu},\label{f1}\\
     \d_\mu\varphi^\mu =-m^2\varphi +\frac{\d\mathcal{L}_\mathrm{int}
     (\varphi,\varphi^\mu)}{\d\varphi},\label{f2}
\end{gather}
which are equivalent to (\ref{f3}). We see that precisely in the case
when the interaction Lagrangian depends on $\d^\mu\varphi$
the interacting field $\varphi^\mu$ differs from $\d^\mu\varphi$.
\bigskip

\noindent {\it Example:} 
By explicit calculation we are going to show that the retarded
products $R_{\mathcal{L}_0}(\varphi^\nu (y),\d^\mu\varphi (x))$ 
and $R_{\mathcal{L}_0}(\varphi^\nu (y),\varphi^\mu(x))$ are different, 
although $\d^\mu\varphi_{\mathcal{L}_0}=\varphi^\mu_{\mathcal{L}_0}$
(where $\mathcal{L}_0$ is the free part of the Lagrangian (\ref{L2})),
so we see again that the entries of the retarded products 
must not be
replaced by their restriction to the space of solutions. The fastest way
to compute these retarded products is to use Proposition 
\ref{prop:ret}(a), as in 
(\ref{S'':inv:KG}). However, we find it more instructive to go back 
to Peierls' definition of retarded products (\ref{R()}):
by definition $r_{\mathcal{L}_0+\lambda\delta_y \varphi^\nu,
\mathcal{L}_0}(f_0,h_0)$ is the solution $(f,h)$ of (\ref{f1})-(\ref{f2})
with $\mathcal{L}_\mathrm{int}(z)=\lambda\delta (z-y)\varphi^\nu (z)$
which agrees in the distant past with $(f_0,h_0)$. We obtain
\begin{gather}
    \varphi(x)\circ r_{\mathcal{L}_0+\lambda\delta_y \varphi^\nu,
    \mathcal{L}_0}(f_0,h_0)=f(x)=
    f_0(x)-\lambda\d^\nu\Delta^\mathrm{ret}(x-y),\notag\\
    \varphi^\mu (x)\circ r_{\mathcal{L}_0+\lambda\delta_y \varphi^\nu,
    \mathcal{L}_0}(f_0,h_0)=h^\mu(x)\notag\\
    =h_0^\mu(x)-\lambda (\d^\mu\d^\nu\Delta^\mathrm{ret}(x-y)-
    g^{\mu\nu}\delta (x-y)).
\end{gather}
Hence,
\begin{equation}
    R_{\mathcal{L}_0}(\varphi^\nu (y),\d^\mu\varphi (x))=
    \d^\mu_xR_{\mathcal{L}_0}(\varphi^\nu (y),\varphi (x))=
    -\d^\nu\d^\mu\Delta^\mathrm{ret}(x-y),
\end{equation}
but    
\begin{equation} 
    R_{\mathcal{L}_0}(\varphi^\nu (y),\varphi^\mu(x))=
    -(\d^\nu\d^\mu\Delta^\mathrm{ret}(x-y)-g^{\nu\mu}\delta (x-y)).
    \label{R=Delta1}
\end{equation}
\section{The Master Ward Identity}
\subsection{Generalized Schwinger-Dyson Equation}
It is an immediate consequence of the field equations $\Bigl(
\frac{\delta S}{\delta\varphi}\Bigr)_S=0$ for a given 
local action $S$
that all functionals of the form 
\begin{equation}
        \Bigl(A\frac{\delta S}{\delta\varphi}\Bigr)(h)
        \label{GSD}
\end{equation}
(by which we mean the point-wise product of an arbitrary 
classical field $A\in\mathcal{P}$
with $\frac{\delta S}{\delta\varphi}$ smeared out with the test
function $h$) vanish on the space of solutions 
$\mathcal{C}_{\mathcal{S}}$.

If we set $S=S_0+\lambda S_1$,  
and differentiate with
respect to $\lambda$ at $\lambda =0$ we obtain the identity
\begin{equation}
    R_{S_0}\Bigl(S_1,A\frac{\delta S_0}{\delta\varphi}(h)
   \Bigr)+\Bigl(A\frac{\delta S_1}{\delta\varphi}(h)
   \Bigr)_{S_0}=0.
   \label{GSchDy}
\end{equation}
For $A\equiv 1$ this equation looks similar to the Schwinger-Dyson
equation
\begin{equation}
    \frac{i}{\hbar}\langle TS_1\frac{\delta S_0}{\delta\varphi}\rangle +
    \langle \frac{\delta S_1}{\delta\varphi}\rangle =0,
    \label{SchDy}
\end{equation}
which holds for the vacuum expectation values of time ordered products
for a quantum field theory with action $S_0$ (see, e.g. \cite{HT}). 
(Note that the
factor $\frac{i}{\hbar}$ is absorbed in (\ref{GSchDy}) in the retarded
part of the Poisson bracket.) For this reason we call 
(\ref{GSchDy})
the {\bf retarded Schwinger-Dyson Equation} and the vanishing of 
(\ref{GSD}) the {\bf generalized Schwinger-Dyson Equation} (GSDE). 
Note that the 
retarded Schwinger-Dyson Equation has the same form in 
classical physics as in quantum physics.

In the retarded Schwinger-Dyson equation we may  
permute the two entries in the retarded product. Namely, 
the difference is just the Poisson bracket which vanishes 
if one of the entries vanishes on the space of solutions.
 
For the retarded products of local fields we obtain the perturbative 
version of the generalized Schwinger-Dyson Equation,
\begin{multline}
    R_{n,1}\bigl(f_{1},\ldots,f_{n},h\frac{\delta 
    S_0}{\delta\varphi}\bigr)+\\
    \sum_{l=1}^n R_{n-1,1}  
    \bigl(f_{1},\ldots,f_{l-1},f_{l+1},\ldots,f_{n},h
    \frac{\delta f_{l}}{\delta\varphi}\bigr)=0
    \label{GSchDy:R}
\end{multline}
with $f_{i},h\in\mathcal{D}(\MM,\mathcal{P}),i=1,\ldots,n$, 
and
where the functional derivative of $f\in\mathcal{D}(\MM,\mathcal{P})$ 
is defined by
\begin{equation}
        \frac{\delta f}{\delta\varphi}(x)=\frac{\delta 
        \int dy\,f(y)}{\delta\varphi(x)}\>,
        \label{eq:functionalderivativeoffields}
\end{equation}
i.e.
\begin{equation}
    \frac{\delta 
    f}{\delta\varphi}=\sum_{a}(-1)^{|a|}\d^{a}\frac{\d 
    f}{\d(\d^{a}\varphi)} \ .   
    \label{eq:functionalderivativeoffields1}
\end{equation}

Proceeding by induction on $n$ and  
using the GLZ-Relation (\ref{Pb(R)}), we obtain an equivalent 
formula for the case that the term $h\frac{\delta 
S_{0}}{\delta\varphi}$ is one of the first $n$ entries of 
$R_{n,1}$, namely
\begin{multline}
        R_{n,1}\bigl(f_{1},\ldots,f_{n-1},h\frac{\delta 
    S_0}{\delta\varphi},f_{n}\bigr)+\\
    \sum_{l=1}^n R_{n-1,1}  
    \bigl(f_{1}, \ldots,f_{l-1}, 
    h\frac{\delta f_{l}}{\delta\varphi}, 
    f_{l+1},\ldots,f_{n}
    \bigr)=0.
        \label{eq:GSchDy:R'}
\end{multline}

The equations (\ref{GSchDy:R}) and (\ref{eq:GSchDy:R'}) remain 
meaningful in perturbative QFT. Also there they are 
equivalent since the GLZ-Relation still holds. We require 
either of them
as a renormalization condition (see Sect.~4). 
Equivalently, an analogous identity may be postulated for 
the time ordered products (where the two versions above coincide in view of 
the symmetry of time ordered products). It is a 
generalization of the condition (N4) in \cite{DF}.\footnote{A first
  generalization (in the framework of causal perturbation theory)
  of the renormalization condition (N4) was given in an
  unpublished preversion of \cite{Pinter}.} But 
there, following the tradition in causal perturbation 
theory, we considered the entries of time ordered or retarded 
products as Wick polynomials of the free field. Therefore, 
time ordering and partial derivatives did not commute, and the 
formulation of identities involving derivatives of fields 
contained many free parameters. A consistent choice of these 
parameters was made possible by the MWI proposed in \cite{DF3}.
  
The QFT version of (\ref{GSchDy:R}) or (\ref{eq:GSchDy:R'}) seems to
correspond to the 'broomstick identity' of Lam given in Fig.~8
of \cite{Lam2}. We think that Lam is unable to write down this
identity as an 
{\it equation} because the arguments of his time-ordered products are
on shell fields; and compared to \cite{DF3} (which uses also an on
shell formalism, cf. Sect.~4) he is not equipped with the 
'external derivative'.

We will see that the MWI is equivalent to the 
generalized Schwinger-Dyson Equation (\ref{GSchDy:R}). 
Therefore, the {\it MWI 
can be interpreted as a quantum version of all identities 
for local fields which 
follow in classical field theory from the field equations}.
\subsection{Definition of a map $\sigma$ from free fields to 
unrestricted fields}
To keep the formulas simple, we consider the case of one real scalar 
field $\varphi$. The procedure, however, applies to a general model. 
Let $\mathcal{J}$ denote the ideal in the algebra $\mathcal{P}$ 
of polynomials of fields which 
is generated from the free 
field equation,
\begin{equation}
        \mathcal{J}=\{\sum_a A_{a}\d^{a}(\square +m^2)\varphi,\, 
        A_{a}\in\mathcal{P},a\in\NN_{0}^{d}\} \ ,
        \label{eq:ideal of free fields}
\end{equation}
let $\mathcal{P}_{0}=\mathcal{P}/\mathcal{J}$ be the algebra 
of free fields and let $\pi:\mathcal{P}\to\mathcal{P}_{0}$ 
be the canonical surjection. Since $\mathcal{J}$ is 
translation invariant, we may define derivatives with 
respect to space-time coordinates in $\mathcal{P}_{0}$ by  
$\d^\mu\pi(B)\=d\pi(\d^\mu B)$, and in this sense
the free field equation holds true for $\pi\varphi$.

The wanted map $\sigma$ is a section
$\sigma :\mathcal{P}_0\rightarrow\mathcal{P}$. In contrast to the
surjection $\pi$, the section $\sigma$ is not canonically given.
We restrict its choice by the following requirements:
\begin{enumerate}
    
     \item  $\pi\circ \sigma=\mathrm{id}$, i.e. $\sigma$ is a section.

     \item  $\sigma$ is an algebra homomorphism.\footnote{In case 
     of complex 
     fields we additionally require $\sigma (A^*)=\sigma (A)^*$.}
     
     \item  The Lorentz transformations commute with $\sigma\pi$.

     \item   $\sigma\pi (\mathcal{P}_1)\subset \mathcal{P}_1$, where
     $\mathcal{P}_1$ is the subspace  of fields
     which are linear in $\d^a\varphi$.

     \item  $\sigma\pi$ does not increase the mass dimension of the fields,
     i.e. $\sigma\pi (B)$ is a sum of terms with mass dimension 
     $\leq\mathrm{dim}\>(B)$. 
     In particular we find $\sigma\pi(\varphi)=\varphi$.

     \item  $\mathcal{P}$ is generated by fields in the image 
     of $\sigma$ and their derivatives. In the present case 
     (one real scalar field), this condition is automatically satisfied. 

\end{enumerate}

By (i) $\sigma\pi:\mathcal{P}\rightarrow\mathcal{P}$ is a projection: 
$\sigma\pi \sigma\pi=\sigma\pi$. The linearity of $\sigma$ and condition (i) 
imply $\mathrm{ker}\>\sigma\pi=\mathrm{ker}\>\pi=J$, and hence
\begin{equation}
    \sigma\pi\square \varphi=-m^2\sigma\pi\varphi \ .
    \label{spi:ffeq}
\end{equation}
We are now looking for the most general explicit formula for $\sigma\pi$
which satisfies the above requirements. Due to (ii) it suffices to
determine $\sigma\pi (\d^a\varphi)$. 
By definition of $\pi$ and $\sigma$ it must hold
\begin{equation}
   \sigma\pi (A)-A\in J\quad\quad \forall A\in\mathcal{P}
   \label{spiA-A}
\end{equation}
and hence
\begin{equation}
    \sigma\pi\d^{a}\varphi = \d^{a}\varphi+ \sum_b c^{a}_{b}\d^{b}(\square 
    +m^2)\varphi 
   \label{spi:ansatz}
\end{equation}
with constants
$c^{a}_{b}\in\RR$.

The determination of an admissible section $\sigma$ satisfying 
conditions (i) to (v) is 
equivalent to the determination of a complementary 
subspace $K=\sigma\pi(\mathcal{P}_{1})$ of $\mathcal{J}_{1}\=d 
\mathcal{J}\cap \mathcal{P}_{1}$ which is Lorentz invariant 
and satisfies the condition that already the subspaces with 
mass dimension $\le n$ are complementary,
\begin{equation}
      K^{(n)}+\mathcal{J}_{1}^{(n)}=\mathcal{P}_{1}^{(n)}\ .
      \label{eq:complement}
\end{equation}
Since the finite dimensional representations of the Lorentz 
group are completely reducible, this is always possible. 
Namely, for the lowest mass dimension $n_{0}=(d-2)/2$ the subspace 
generated by the field equation is zero, thus 
$K^{(n_{0})}=\mathcal{P}_{1}^{(n_{0})}=\RR \varphi$. From this 
the existence of $K^{(n+1)}$ may be proved by induction on $n$. One 
just has to choose a Lorentz invariant complementary 
subspace $L^{(n+1)}$ of $K^{(n)}+\mathcal{J}_{1}^{(n+1)}$ 
in $\mathcal{P}_{1}^{(n+1)}$ and to set 
$K^{(n+1)}=K^{(n)}+L^{(n+1)}$.

The arbitrariness in the choice of $L^{(n)}$ depends on the 
multiplicity in which the irreducible subrepresentations of 
the Lorentz group occur in the respective subspaces. 
In the present case it turns out that $\sigma$ is unique 
(see the second part of Appendix A).  

In case one introduces the auxiliary field 
$\varphi^{\mu}$, the choice of $\sigma$ involves free 
parameters.
A special choice for $\sigma$ is given in the first part of Appendix A.   
For the lowest derivatives we obtain the
following general solution of the requirements (i)-(vi):
\begin{gather}
   \sigma\pi (\varphi) =\varphi,
   \label{spi:0}\\
   \sigma\pi (\d^\mu\varphi)=
   \sigma\pi (\varphi^\mu)=\gamma\varphi^\mu +
   (1-\gamma)\d^\mu\varphi ,\quad \gamma\in\RR\setminus\{0\},
   \label{spi:1}\\
   \sigma\pi (\d^\mu\d^\nu\varphi)=\sigma\pi (\d^\nu\varphi^\mu)=
   (1+2\alpha)\d^\mu\d^\nu\varphi \notag\\
   -\alpha (\d^\mu\varphi^\nu +
    \d^\nu\varphi^\mu)-\frac{1+2\alpha}{d}g^{\mu\nu}\square\varphi
   -\frac{1}{d}g^{\mu\nu}m^2\varphi +\frac{2\alpha}{d}g^{\mu\nu}
   \d^\sigma\varphi_\sigma,
   \label{spi:2}
\end{gather}
where $\gamma$ and $\alpha\in\RR$ are free parameters. The 
condition $\gamma\ne 0$ is necessary and sufficient for (vi)
provided (i)-(v) are satisfied. A preferred
choice is $\gamma =1$. 

\subsection{The Master Ward identity}
Let $A$ be a functional of the field which vanishes according to 
the field equation derived from the action $S_0$, i.e. 
$A_{S_0}=0$. $A$ is of the form (cf. the remark in footnote 3)
\begin{equation}
    A=\int dx\,G(x)\frac{\delta S_0}{\delta \varphi(x)}\ ,
\end{equation}
with $G(x)\in \mathcal{F}(\mathcal{C})$. (This formula states that $A$
is an arbitrary element of the ideal ${\mathcal J}_{S_0}$ generated 
by the field equation belonging to $S_0$.)
We may introduce the derivation
\begin{equation}
    \delta_A=\int dx\,G(x)\frac{\delta}{\delta \varphi(x)}
\end{equation}
on $\mathcal{F}(\mathcal{C})$. The GSDE imply
\begin{equation}
    \mathbf{(MWI)}\quad\quad\quad
    R_{S_0}(e_{\otimes}^{S_1},A)=-R_{S_0}(e_{\otimes}^{S_1},\delta_A(S_1))  
    \quad\forall S_1\in \mathcal{F}(\mathcal{C})\label{mwi:R1}
\end{equation}
and, by using the GLZ equation
\begin{equation}
    R_{S_0}(e_{\otimes}^{S_1}\otimes A,B)=
    -R_{S_0}(e_{\otimes}^{S_1}\otimes\delta_A(S_1),B)
    -R_{S_0}(e_{\otimes}^{S_1},\delta_A(B))
    \quad \forall S_1,B\in \mathcal{F}(\mathcal{C}) \ .\label{mwi:R2}
\end{equation}
This equation (in a different form, see (\ref{mwi}) below) 
was proposed by Boas and D\"utsch \cite{DF3} under the 
name {\bf Master Ward Identity} (MWI) as a universal 
renormalization condition in perturbative QFT. 
In the present formulation it is evidently 
equivalent to the GSDE. Note that up to now (in contrast to \cite{DF3}) 
the formulation of the MWI makes sense also in the case that $S_0$ is not 
a quadratic functional of the field. Similarly to the GSDE, the MWI
holds also non-perturbatively:
\begin{equation}
    A_{S_0+S_1}=-\bigl(\delta_A(S_1)\bigr)_{S_0+S_1},\quad A\in
    {\mathcal J}_{S_0},\> S_1\in \mathcal{F}(\mathcal{C}).\label{mwi:neu}
\end{equation}
\bigskip

\noindent {\it Example:} 
As a typical application let us look at the free complex scalar field 
with the conserved current
\begin{equation}
    j_{\mu}=\frac{1}{i}(\varphi^*\d_{\mu}\varphi -\varphi\d_{\mu}\varphi^*) \ .
\end{equation}
Let $A=\langle\d j,g\rangle \equiv \int dx\,\d^{\mu}j_{\mu}(x)g(x)$ with 
$g\in\mathcal{D}(\MM)$. We have
\begin{equation}
     A=\frac{1}{i}
     \left(
       \langle g\varphi^*,\frac{\delta S_0}{\delta \varphi^*}\rangle -
       \langle g\varphi,\frac{\delta S_0}{\delta \varphi}\rangle 
     \right)
\end{equation}
and hence
\begin{equation}
     \delta_A=\frac{1}{i}
     \left(
       \langle g\varphi^*,\frac{\delta}{\delta \varphi^*}\rangle -
       \langle g\varphi,\frac{\delta}{\delta \varphi}\rangle 
     \right)\ .
\end{equation}
If $g\equiv 1$ on the localization region of $F\in \mathcal{F}(\mathcal{C})$,
$\delta_A F$ is the infinitesimal gauge transformation of $F$, and inserting 
$A$ into the MWI yields the well known Ward identity of the model.
\bigskip

A problem with the MWI in the form presented above is the 
non-uniqueness of the derivation $\delta_A$ for a given $A$,
e.g. for 
\begin{equation}
A=\int dx\, h(x)((\square +m^2)\varphi^* (x))
(\square +m^2)\varphi (x) 
\end{equation} 
in case of the complex scalar field.
We therefore turn now to the free field case and use the 
techniques and conventions developed in the preceding section.

We will give a unique prescription to write any
$A=\int dx\,h(x)B(x)$ with $B\in \mathcal{J}\subset\mathcal{P}$ and
$h\in\mathcal{D(\MM)}$ (i.e. $A\in \mathcal{J}_{S_0}$) 
in the form $A =\int dx\,h(x)\sum_j B_j(x)b_j(x)$ with $b_j\in
\mathcal{P}_1\cap\mathcal{J}$.
Then we may set
\begin{equation}
  \delta_A =\int dx\,h(x)\sum_j
   B_j(x)\delta_{b_j(x)}
\end{equation} 
with
\begin{equation}
   \delta_{b_j(x)}(F)=-R_{S_0}(b_j(x),F) \ , \quad  
   F\in\mathcal{F}(\mathcal{C})\ ,
\end{equation}   
where we used (\ref{a30}) and the fact that for terms which are 
linear in the field the first term on the right hand side vanishes, 
such that (\ref{a31}) 
holds everywhere, not only on the 
space of solutions.

To give the mentioned prescription let $B\in \mathcal{J}$. 
Then $B=p(B)$ where $p=1-\sigma\pi$ is 
a projection from $\mathcal{P}$ onto $\mathcal{J}$. Since $\sigma\pi$
is an algebraic homomorphism we find
\begin{equation}
   p(B_1B_2)=B_1p(B_2)+p(B_1)\sigma\pi(B_2)\ . 
\end{equation}
Hence, we may write every $B\in \mathcal{J}$ in the form
\begin{equation}
   B=\sum_{\chi\in\mathcal{G}}B_{\chi}p(\chi) 
\end{equation}
where we introduced the set $\mathcal{G}$ 
of generators  of the field algebra
$\mathcal{P}$,
\begin{equation}
    \mathcal{G}\=d\{\d^a\varphi\> |\> a\in\NN_0^d\}
\end{equation}
which is a vector space basis of $\mathcal{P}_1$. The coefficients $B_{\chi}$
can be found in the following way:
Any $B\in \mathcal{P}$ is a polynomial $P$ in finitely many different elements 
$\chi_1,\ldots,\chi_n \in \mathcal{G}$, 
$B=P(\vec{\chi}), \vec{\chi}=(\chi_1,\ldots,\chi_n)$.   
Since $B\in \mathcal{J}$ we have
\begin{gather}
   B=B-\sigma\pi(B)=P(\vec{\chi})-P(\sigma\pi(\vec{\chi}))\\
   =\int_0^1 d\lambda \frac{d}{d\lambda} 
   P(\lambda \vec{\chi}+ (1-\lambda)\sigma\pi(\vec{\chi}))\\
   =\sum_{i=1}^n P_i(\vec{\chi},\sigma\pi(\vec{\chi}))p(\chi_i)
\end{gather}
with 
\begin{equation}
   P_i(\vec{\chi},\sigma\pi(\vec{\chi}))=
  \int_0^1 d\lambda 
  \d_i P(\lambda \vec{\chi}+ (1-\lambda)\sigma\pi(\vec{\chi})).    
\end{equation}
Hence we may set $B_{\chi_i}=P_i(\vec{\chi},\sigma\pi(\vec{\chi}))$, 
$i=1,\ldots,n$ and $B_{\chi}=0$ if $\chi\not\in \{\chi_1,\ldots\chi_n\}$.
\bigskip

\noindent {\it Example:} 
Let $\varphi$ be a real scalar field. For $\chi
=\varphi,\>\d^\mu\varphi$
the expression $p(\chi)$ vanishes. But for 
$\chi=\d^\mu\d^{\nu}\varphi$ we obtain
$p(\chi)=\frac{g^{\mu\nu}}{d}(\square +m^2)\varphi=
\frac{g^{\mu\nu}}{d}\frac{\delta S_0}{\delta\varphi}$
and hence
\begin{equation}
  \delta_{p(\chi)(x)}(F)=\frac{g^{\mu\nu}}{d}\frac{\delta F}{\delta\varphi}\ .
\end{equation}  
\bigskip

In many applications one wants to compare derivatives in the free 
theory with those in the interacting theory. One is therefore 
interested in expressions of the form  
\begin{equation}
  A=\int dx\,h(x)(\d_{\mu}\sigma\pi(B)(x)-\sigma\pi(\d_{\mu}B)(x))
\end{equation}  
with $B\in\mathcal{P}$ and $h\in\mathcal{D}(\MM)$. To get a more
general identity we even admit $h\in\mathcal{D}(\MM,\mathcal{P})$.
Clearly, 
$[\d_{\mu},\sigma\pi](B)\in \mathcal{J}$, hence $A\in\mathcal{J}_{S_0}$. 
Using the Leibniz' rule $\d^\mu B=\sum_{\chi\in\mathcal{G}}
\frac{\d B}{\d\chi}\d_{\mu}\chi$ we find
\begin{equation}
  [\d_{\mu},\sigma\pi](B)=
  \sum_{\chi\in\mathcal{G}}\sigma\pi(\frac{\d B}{\d\chi})
  [\d_{\mu},\sigma\pi](\chi) 
\end{equation}  
and get the formula
\begin{equation}
 \delta_A =\int dx\,h(x)\sum_{\chi\in\mathcal{G}}
   \sigma\pi(\frac{\d B}{\d\chi})(x)\delta_{[\d_\mu,\sigma\pi](\chi)(x)}\ .
\end{equation}  
In terms of the retarded fields (\ref{fi^ret}) we end up with
   \begin{equation}
        \mathbf{(MWI')}\quad\quad\quad
        \bigl( h([\d^\mu,\sigma\pi] B) 
        \bigr)_{\mathcal{L}_\mathrm{int}}=
        \sum_{\chi,\psi\in {\cal G}}
        \Bigl(h\>\sigma\pi
        \Bigl(\frac{\d B}{\d\chi}\Bigr)\>
        \delta^{\mu}_{\chi,\psi}
        \frac{\d \mathcal{L}_\mathrm{int}}{\d\psi}
        \Bigr)_{\mathcal{L}_\mathrm{int}},
        \label{mwi}
   \end{equation}
where the differential operator $\delta^{\mu}_{\chi,\psi}$ is defined by
\begin{equation}
        \delta_{\chi,\psi}^{\mu}f(x)=-\int dy\, 
        R_{S_0}\bigl([\d^{\mu},\sigma\pi](\chi)(x),\psi(y)\bigr)f(y)\ ,\ 
        f\in\mathcal{D}(\MM,\mathcal{P})\ .
        \label{eq:mwi-preparation4}
\end{equation}
Note that $R_{S_0}\bigl([\d^{\mu},\sigma\pi](\chi)(x),\psi(y)\bigr)$
is a linear combination of partial derivatives of $\delta (x-y)$.
\bigskip

\noindent {\it Example:} 
Let $\chi=\d^{\nu}\varphi$. Then 
$[\d^\mu,\sigma\pi](\chi)=\frac{g^{\mu\nu}}{d}(\square +m^2)\varphi$.
Hence 
$\delta_{[\d^\mu,\sigma\pi](\chi)}=
\frac{g^{\mu\nu}}{d}\frac{\delta}{\delta\varphi}$, 
therefore one obtains for the difference between 
derivatives of free or interacting fields
\begin{equation}
   \d^{\mu}(\d^{\nu}\varphi)_{S_0+S_1}-
   (\sigma\pi\d^{\mu}\d^{\nu}\varphi))_{S_0+S_1}=
   \frac{g^{\mu\nu}}{d}
\left(\frac{\delta S_1}{\delta\varphi}\right)_{S_0+S_1} \ .\label{ex-MWI'}
\end{equation}
We think that the non-vanishing of $[\d^\mu,\sigma\pi](\d^\nu\varphi)$
compared to $[\d^\mu,\sigma\pi](\varphi)=0$ explains why the formalism
of Lam becomes inconsistent for vertices containing higher than first
derivatives of the basic fields (cf. Sect.~V of \cite{Lam1}).
\bigskip

We derived (\ref{mwi}) as a consequence of the 
MWI (\ref{mwi:neu}). It is even equivalent 
if $\mathcal{J}\cap\mathcal{P}_1$ is 
spanned by fields of the form $[\d^{\mu},\sigma\pi](\psi)$ 
and their derivatives, with $\psi\in\mathcal{P}_1$, since 
then the l.h.s. of the MWI (\ref{mwi:R1}) can be written
as a linear combination of fields of the form of the
l.h.s. of (\ref{mwi}). In the case of the free scalar 
field this condition is clearly fulfilled, since 
$[\d^{\mu},\sigma\pi](\d_{\mu}\varphi)=(\square 
+m^2)\varphi$. In the enlarged model with the auxiliary 
field $\varphi_{\mu}$ we find 
$[\d^{\mu},\sigma\pi](\varphi)=\gamma(\d^{\mu}\varphi 
-\varphi^{\mu})$ and $[\d^{\mu},\sigma\pi](\varphi_{\mu})= 
(\d^{\mu}\varphi_{\mu}+m^2\varphi) + 
(\gamma-1)\d^{\mu}(\varphi_{\mu}-\d_{\mu}\varphi)$, hence 
here it follows from $\gamma\ne 0$. Also in general it 
follows from condition (vi) on $\sigma$ in Sect.~3.2. Namely, 
let $\chi\in\mathcal{J}\cap\mathcal{P}_1$. According to (vi) and (iv), 
$\chi$ is of the form 
\begin{displaymath}
        \chi =\sum_{k,a} \d^{a}\sigma\pi (\psi_{k,a})\quad
 \psi_{k,a}\in\mathcal{P}_1\ .
\end{displaymath} 
Hence,
\begin{displaymath}
      \chi =\chi -\sigma(\pi\chi) =  \sum_{k,a} [\d^{a},\sigma\pi]
            \sigma\pi (\psi_{k,a}) \ .
\end{displaymath}    
The result now follows from the derivation property of the commutator 
\begin{displaymath}
        [\d_{\mu_{1}}\cdots\d_{\mu_{n}},\sigma\pi]=\sum_k
    \d_{\mu_{1}}\cdots \d_{\mu_{k-1}}[\d_{\mu_{k}},\sigma\pi] 
    \d_{\mu_{k+1}}\cdots \d_{\mu_{n}} \ .
\end{displaymath}

We point out that the MWI (in the original form (\ref{mwi:R1})
or (\ref{mwi:R2}) as well as in the second form (\ref{mwi})) is well 
defined in perturbative QFT, too. This is a main ingredient of the 
next Section. 
\section{Quantization: defining properties of perturbative quantum 
fields}
The structure of perturbative classical field theory which was 
analyzed in this paper can to a large degree be preserved during 
quantization. The main change is the replacement of the 
commutative product of functionals 
$F\in\mathcal{F}(\mathcal{C})$ by a $\hbar$-dependent 
non-commutative associative product, and by the replacement 
of the Poisson bracket by $\frac{1}{i\hbar}$ times the commutator. 
The definition of the product can be read off from Wick's Theorem,
\begin{equation}
  \label{eq:Wick}
  F*_{\hbar}G=\sum_{n=0}^{\infty}\frac{\hbar^n}{n!}\int d(x,y) 
  \frac{\delta^n F}{\delta\varphi(x_1)\cdots \delta\varphi(x_n)} 
  \prod_{i=1}^n \Delta_{+}(x_i-y_i) 
  \frac{\delta^n G}{\delta\varphi(y_1)\cdots \delta\varphi(y_n)} 
\end{equation}
where $\Delta_+$ denotes the positive frequency fundamental 
solution of the Klein Gordon equation. This abstract algebra 
(which we still denote by $\mathcal{F(C)}$) can be
represented on Fock space by Wick polynomials
\begin{equation}
  \pi(F)=\sum_n\frac{1}{n!}\int d^nx\,
\frac{\delta^n F}{\delta\varphi(x_1)...\delta\varphi(x_n)}
\vert_{\varphi=0}:\varphi(x_1)...\varphi(x_n):\>,
\end{equation}
the kernel of this representation being the set of functionals $F$
vanishing on ${\cal C}_{{\cal L}_0}$, cf. \cite{DF1}.

A direct construction of solutions of the field equations in the 
case of local interactions is, in general, not possible because of 
ultraviolet divergences. One may, however, start from the ansatz
\begin{equation}
  \label{eq:quantum field}
  B_{\mathcal{L}_{\mathrm{int}}}(f)= 
  \sum_{n=0}^{\infty} \frac{1}{n!}
  R_{n,1}(\mathcal{L}_{\mathrm{int}}^{\otimes n},Bf)
\end{equation}
in analogy to (\ref{eq:pert class field})
and try to determine the retarded products $R_{n,1}$ as polynomials in 
$\hbar$ such that they satisfy the following properties:
They are $(n+1)$-linear continuous functionals 
on $\mathcal{D}(\MM,\mathcal{P})$ 
with values in $\mathcal{F}(\mathcal{C})$ which are symmetric 
in the first $n$ entries, have retarded support (\ref{supp(R)}) and 
satisfy the GLZ-Equation (\ref{Pb(R)}). Moreover they have to fulfil 
the unitarity condition
\begin{equation}
   R_{n,1}(f_1\otimes ...\otimes f_n,f)^*=
   R_{n,1}(f_1^*\otimes ...\otimes f_n^*,f^*).
   \label{eq:unitarity}
\end{equation} 

It turns out, that already by these properties, the retarded 
products $R_{n,1}$ are uniquely determined outside of the total 
diagonal $x_1=\ldots=x_{n+1}$ in terms of the lower order retarded products
where the lowest order is defined by $R_{0,1}(Bf)=
B(f)\vert_{\mathcal{C}_{\mathcal{L}_0}}$.
{\it Renormalization then means the extension of the retarded products 
to the diagonal}. This is a variant of the Bogoliubov-Epstein-Glaser 
renormalization method and, in the adiabatic limit $g\equiv 1$, it 
has been worked out by Steinmann \cite{SD}. A modernized 
and local version of the 
procedure will be presented in \cite{DF4}.

The main work which remains to be done is the so-called finite 
renormalization, i.e.,  
the analysis of the ambiguities in the extension process. 

In a first step, the condition (\ref{eq:N3class}) (condition (N3) 
of \cite{DF}) can be used to reduce the extension problem to a problem for 
numerical distributions. By requiring translation invariance these numerical 
distributions depend only on the differences of coordinates, thus one has 
to study the mathematical problem of extending a distribution which is 
defined outside of the origin to an everywhere defined distribution. 
The possible extensions can be classified in terms of Steinmann's \cite{SD}
scaling 
degree, and it is a natural requirement that the scaling degree should 
not increase during the extension process.
In addition one can show that the extension can always be done such 
that the retarded products are Lorentz covariant. 

The steps described above can always be performed and leave, for every 
numerical distribution, a finite set of 
parameters undefined. The proposal is now to add two further conditions.

One is the Action Ward Identity (\ref{[d,R]}) proposed by 
Stora \cite{AWI}. A proof that it can always be satisfied and is 
compatible with the other normalization conditions will be given in
\cite{DF4}. We require then the Master Ward Identity in the form 
(\ref{mwi:R1}) or (\ref{mwi:R2}), or, equivalently, the generalized 
Schwinger-Dyson equation in the form (\ref{GSchDy:R}) or (\ref{eq:GSchDy:R'}).
Here anomalies may occur, and one has to check in a given model whether 
these identities can be satisfied. Fortunately, for a typical
application, one needs these identities only for special cases which
may be characterized by the polynomial degree of the fields and the
number of derivatives which are involved.
\bigskip

The formalism given here is not completely 
equivalent to the one of \cite{DF3}. 
The algebras of {\it symbols} ${\cal P}$ and ${\cal P}_0$ given in
\cite{DF3} may be identified with our  algebras ${\cal P}$ and 
${\cal P}_0$ of {\it classical fields}. The main difference is the absence 
of the Action Ward Identity, thus derivatives could not freely be shifted 
from fields to test functions. Therefore, in \cite{DF3}
an 'external derivative' $\tilde\d^\mu$ on ${\cal P}_0$ 
was introduced which generates
new symbols $\tilde\d^a A$ ($A\in {\cal P}_0$, $a\in \NN_0^d $).
The argument of the retarded product of \cite{DF3}
(we denote it here by $\tilde R_{n,1}$) is an element of
\begin{equation}
   (\tilde {\cal P}_0\otimes {\cal D}(\MM))^{\otimes (n+1)}\quad
   {\rm where}\quad \tilde {\cal P}_0\=d \bigvee
   \{\tilde\d^a A\,|\,A\in {\cal P}_0,
   a\in \NN_0^4\}
\end{equation}
(for details see \cite{DF3}). The retarded products
$\tilde R_{n,1}$ of symbols with external derivative(s) can be defined in 
terms of 
retarded products without external derivative 
(normalization condition ${\bf (\tilde N)}$). 
The MWI is then
expressed by a further normalization condition {\bf (N)}, combined
with ${\bf (\tilde N)}$. In particular ${\bf (\tilde N)}$ and
{\bf (N)} imply
\begin{equation}
  \tilde R_{n,1}(W_1g_1,\ldots, (\tilde\d^\nu W_l)g_l+W_l\d^\nu g_l,\dots, 
  W_{n+1}g_{n+1})=0 \ ,
  W_1,...W_{n+1}\in {\cal P}_0 \ .
  \label{R(tilde-d)}
\end{equation}

To clarify the relation of the two formalisms we 
extend $\sigma$ to a map $\tilde\sigma :\tilde{\mathcal{P}}_0
\rightarrow\mathcal{P}$ by setting
\begin{equation}
  \tilde\sigma(\tilde\d^a A)\=d\d^a\sigma (A),\quad A\in\mathcal{P}_0,
  \label{tilde-sigma}
\end{equation}
and requiring that $\tilde\sigma$ is an algebra $*$-homomorphism,
similarly to \cite{DF3}.
The defining property (vi) of $\sigma$ means that $\tilde\sigma$ 
is surjective. But  $\tilde\sigma$ is not injective, even if the 
auxiliary field $\varphi^\mu$ is introduced.
So, from the retarded product $R_{n,1}$
of this paper, we can construct a retarded product $\tilde R_{n,1}$
in the sense of \cite{DF3}, by defining
\begin{equation}
  \tilde R_{n,1}(W_{1}g_1,\ldots,W_{n+1}g_{n+1})
  \=d R_{n,1}(\tilde\sigma(W_{1})g_1,\ldots,\tilde\sigma(W_{n+1})g_{n+1}).
  \label{R>tilde R}
\end{equation}
But, in general, it might happen that $\tilde R_{n,1}$ does not vanish 
if one entry is in $(\mathrm{ker}\>
\tilde\sigma)$, and then it would be impossible to
construct $R_{n,1}$ from $\tilde R_{n,1}$. If the $R_{l,1}, 
\>l\leq n$, satisfy all defining properties given here (including
the AWI and the MWI), then the corresponding $\tilde R_{l,1}, 
\>l\leq n$, (\ref{R>tilde R}) 
fulfill the requirements on a retarded product given in \cite{DF3},
in particular ${\bf (\tilde N)}$ and {\bf (N)}.
 
We see, that the normalization
conditions on $\tilde R_{n,1}$ are weaker than the normalization
conditions on $R_{n,1}$: (\ref{R(tilde-d)}) can always be 
fulfilled by definition (see \cite{DF3}), even for 
$W_1,...,W_{n+1}\in \tilde {\cal P}_0$. On the other side it is 
still unclear whether one can always satisfy the 'Action Ward Identity'. 
It seems, however, {\it that the formalism given here is the natural one when
departing from classical field theory}.\bigskip

\noindent {\it Remark}: 
In the formalism of \cite{DF3}  and in \cite{DHKS} 
the Feynman (or retarded) propagators
of perturbative QFT contain undetermined parameters, if there are at
least two derivatives present (in $d=4$ dimensions). 
On the classical side (retarded) fields
and their perturbative expansion are unique. The non-uniqueness is
located in the choice of the map $\sigma$: {\it the free parameters in 
$\sigma$ can be identified with the free parameters in the
Feynman propagators of QFT}. This is obvious from
\begin{equation}
  \Delta^\mathrm{ret}_{\varphi_0,\chi_0}(x-y)\=d
\tilde R_{1,1}(\chi_0(y);\varphi_0(x))=R_{1,1}(\tilde\sigma(\chi_0)(y),
\tilde\sigma(\varphi_0)(x)).
\end{equation}
In the non-enlarged
formalism (without $\varphi^\mu$), in which $\sigma$ is unique, a
particular choice of the parameters in the Feynman propagators 
of \cite{DF3} is done.
\section{Application to BRS-Symmetry}
\subsection{Motivation}
The canonical formalism as developed in this paper cannot directly be 
applied to gauge theories because there the Cauchy problem is ill posed 
due to the existence of time dependent gauge transformations. As usual,
one may add a gauge fixing term as well as a coupling to ghost and 
antighost fields to the Lagrangian such that the Cauchy problem 
becomes well posed. The algebra of observables is then obtained as the 
cohomology of the BRS transformation $s$ \cite{BRS}
which is a graded derivation 
which is implemented by the BRS charge $Q$. In QFT one finally 
constructs the space of physical states as the cohomology of $Q$
(see e.g. Sects.~4.1-4.2 of \cite{DF}).

The implementation of this program in the case of  
perturbative gauge field theory meets the problem that in general
the BRS operator $Q$ is changed due to the interaction
\cite{DF}. It is a major problem to exhibit 
the corresponding Ward identities which generalize the 
Slavnov Taylor identities to the case of {\it couplings of compact support}.

In the case of a purely massive theory one may adopt a formalism due to 
Kugo and Ojima \cite{KO}
who use the fact that in these theories the BRS charge $Q$ 
can be identified with the incomimg (free) BRS charge,
which we denote by $Q_0$. For the 
$S$-matrix to be a well defined operator on the physical Hilbert 
space of the free theory one then has to require
\begin{equation}
[Q_0,T((gL)^{\otimes n})]|_{\mathrm{ker}Q_0}\to 0\label{[Q,S]=0}
\end{equation}
in the adiabatic limit $g\to 1$, see e.g. \cite{DSchroer,G}. This is
the motivation to require 'perturbative gauge invariance' 
\cite{DHKS,DS,Scharf}, which is a somewhat stronger condition than 
(\ref{[Q,S]=0}) but has the advantage that it is well defined
independent of the adiabtic limit. The condition (\ref{[Q,S]=0})
(or perturbative gauge invariance) can be satisfied if additional scalar 
fields (corresponding to Higgs fields) are included. Unfortunately, in the 
massless case, it is unlikely that the adiabatic limit
exists.\footnote{To 
motivate perturbative gauge invariance in that case one can derive 
it from a suitable form of conservation of the BRS-current.
To lowest orders this results (as a byproduct) from Appendix B.}

So, in the general case an $S$-matrix formalism is problematic. 
One should better rely on the construction of 
local observables in terms of couplings with compact support. But then 
$Q$ is a formal power series with zeroth order term $Q_0$, and it is 
not obvious which conditions one should put on the retarded 
(or time ordered) products. 

The difficulty is that 
one has to formulate symmetry conditions for the perturbed fields 
which themselves are deformed due to the interaction.
But using the formalism of the present 
paper we can disentangle these two problems. Namely, we first use 
the MWI together with the AWI to compute the commutator of the 
free BRS charge $Q_0$ with the retarded (or 
time ordered) products. The resulting 
family of identities is called the Master BRST Identity and may be used as a 
renormalization condition in its own right. One then can formulate 
conditions on the interaction which 
ensure that the Master BRST Identity implies BRS-invariance of the
interacting theory.
\subsection{The free BRS-transformation}
We illustrate the general ideas on the example of
$N$ massless gauge fields $A^\mu_a,\>a=1,...,N$, each of
them accompanied by a pair of fermionic ghost fields 
$\tilde u_a,\,u_a$. We may also introduce auxiliary fields $B_a$ 
(the Nakanishi-Lautrup fields \cite{NL}).
We work in Feynman gauge, in which the free field equations read
\begin{equation}
   \square A^\mu_{a}=0,\quad\quad \square u_{a}=0=
   \square \tilde u_{a},\quad\quad\forall a
\end{equation}
together with the equation for the auxiliary field
\begin{equation}
  \d_{\mu}A_a^{\mu}=B_a \ .
\end{equation}
We omit in what follows the colour index $a$ by using matrix notation.

The free BRS-current 
\begin{equation}
  j^\mu \=d B\d^\mu u- 
(\d^\mu B) u\label{BRS-current}
\end{equation}
is conserved due to the free field equations,
$\d_\mu j^\mu=B\square u - u \square B$.
The corresponding charge
\begin{equation}
  Q_0\=d\left(\int_{x^0={\rm const.}}d^3 x j^0(x)\right)_{S_0},\label{Q_0}
\end{equation}
is nilpotent, i.e. $Q_0^2=0$ and  
\begin{equation}
  \{Q_0,Q_0\}_{S_0}=0 \ ,
\end{equation}
where we introduced a grading into our Poisson bracket corresponding 
to ghost number.

Using current conservation as well as the GLZ relation we find for the 
Poisson bracket of $Q_0$ with a retarded product
\begin{equation}
  \{Q_0,R_{S_0}(F_1,\ldots,F_n)\}_{S_0}=
  -R_{S_0}(\langle\d j,h\rangle,F_1,\ldots,F_n) \ ,\label{[Q_0,R]}
\end{equation}
where $h\equiv 1$ on a causally complete open region $\mathcal{O}$ 
containing the localization regions of all $F_i\in\mathcal{F}(\mathcal{C})$,
$i=1,\ldots,n$ (cf. the analogous argument for time ordered products 
in \cite{DF1}).\footnote{It is instructive to derive (\ref{[Q_0,R]}) 
in terms of retarded products: let
$\partial^{\mu} h=b^{\mu}-a^{\mu}$
with $\supp b^{\mu}\cap (\overline{V}_++{\cal O})=\emptyset$ and  
$\supp a^{\mu}\cap (\overline{V}_-+{\cal O})=\emptyset$. Since
$R_{S_0}(F_1,\ldots,F_n)$ is localized in ${\cal O}$,
we may vary $b^\mu$ in the spacelike 
complement of $\overline{\cal O}$ without affecting $\{\langle j_\mu ,
b^\mu\rangle_{S_0},R_{S_0}(F_1,\ldots,F_n)\}$. In this way and by using 
$(\d^\mu j_\mu (x))_{S_0}=0$ we find
\begin{equation}
\{Q_0,R_{S_0}(F_1,\ldots,F_n)\}_{S_0}=
\{\langle j_\mu ,
b^\mu\rangle_{S_0},R_{S_0}(F_1,\ldots,F_n)\}.\label{QW}
\end{equation}
By means of the support property (\ref{supp(R)}) of the retarded products 
and the GLZ-Relation (\ref{Pb(R)}) we obtain for the r.h.s. of (\ref{QW})
\begin{gather}
  =\sum_{I\subset \{1,...,n-1\}}\{R_{S_0}((F_l)_{l\in I},\langle j ,
b\rangle),R_{S_0}((F_k)_{k\in I^c},F_n)\}\notag\\
=R_{S_0}(\langle j,b\rangle,F_1,\ldots,F_n)-
R_{S_0}(F_1,\ldots,F_n,\langle j,b\rangle)\notag\\
=R_{S_0}(\langle j,b\rangle,F_1,\ldots,F_n)=-
R_{S_0}(\langle\d j,h\rangle,F_1,\ldots,F_n) \ .\label{QW1}
\end{gather}
} To avoid signs which are due to ferminoic permutations, we assume
that all (local) functionals $F_1,...,F_n$ are bosonic, i.e. a field
polynomial with an odd ghost number is smeared out with a Grassmann
valued test function. 

The free field equations are derived from the Lagrangian 
\begin{equation}
  \mathcal{L}_0=\frac{1}{2}\d_{\mu}A_{\nu}(\d^{\nu}A^{\mu}-\d^{\mu}A^{\nu})+
  \d_{\mu}\tilde{u}\,\d^{\mu}u - B \d_{\mu}A^{\mu} +\frac{1}{2}B^2
\end{equation}
hence
\begin{equation}
  \square u=-\frac{\delta S_0}{\delta\tilde{u}} \ ,\ 
  \square B=\d_{\mu}\frac{\delta S_0}{\delta A_{\mu}}
\end{equation}
with $S_0=\int \mathcal{L}_0$.
Thus we obtain
\begin{equation}
  \delta_{\d j(x)}=-B(x)\frac{\delta}{\delta\tilde{u}(x)}- 
  u(x)\d_{\mu}\frac{\delta}{\delta A_{\mu}(x)}\=d \tilde s_0(x) \ ,
  \label{s_0(x)}
\end{equation}
hence 
\begin{equation}
   \delta_{\langle\d j,h\rangle}=\int dx\,h(x)\tilde s_0(x)\=d s_0\label{s_0}
\end{equation}
on fields localized in the region where $h\equiv 1$, i.e. we obtain 
the free BRS transformation
\begin{equation}
   s_0(\tilde{u})=-B\ ,\ s_0(B)=0\ ,\ s_0(A_{\mu})=\d_{\mu}u\ ,\ s_0(u)=0\ ,
\end{equation}
which is obviously nilpotent. Note that the 'local' free
BRS-transformation $\tilde s_0(x)\int dy\, f(y)$ (where
$f\in\mathcal{P}\otimes\mathcal{D}(\MM)$) differs from $s_0 f(x)$
by a sum of divergences. We now apply the MWI to 
(\ref{[Q_0,R]}) and find the identity 
\begin{equation}
 \{Q_0,R_{S_0}(F_1,\ldots,F_n)\}_{S_0}=
  \sum_{k=1}^n 
R_{S_0}(F_1,\ldots,s_0(F_k),\ldots,F_n) \ .\label{mbrst:s_0}
\end{equation}
In \cite{DF3} this identity (in a somewhat different form, see 
(\ref{mbrst:G^1}) and (\ref{mbrst:G}) below) is called {\bf'Master BRST 
Identity'}.

We may ask how the Master BRST Identity (\ref{mbrst:s_0})
changes if one eliminates the 
Nakanishi-Lautrup field $B$ by using the field equation $B=\d A$. The 
problem is that the ideal $\mathcal{J}_B$ generated from $B-\d A$ in the 
algebra $\mathcal{P}_B$ of all polynomials in the fields and their derivatives
is not stable under $s_0$. We discuss two possibilities. 
\begin{itemize}

\item The quotient algebra 
$\mathcal{P}=\mathcal{P}_B/\mathcal{J}_B$ may be identified with the 
subalgebra of polynomials in $\mathcal{P}_{B}$ which do not contain $B$ 
or its derivatives. Let $\sigma_B:\mathcal{P}\to\mathcal{P}_B$ denote 
this identification (i.e. $\sigma_B(\d^aB+\mathcal{J}_B)=\d^a\d_\nu A^\nu$)
and $\pi_B:\mathcal{P}_B\to\mathcal{P}:X\rightarrow X+\mathcal{J}_B$ 
the canonical homomorphism. Then we set
\begin{equation}
  t=\pi_B s_0\sigma_B 
\end{equation}
i.e.
\begin{equation}
  t(\tilde{u})=\d A \ ,\ t(A_{\mu})=\d_{\mu}u \ , t(u)=0 \ .
\end{equation}
This choice has the disadvantage that $t^2\neq 0$. On the other hand, 
it has the advantage that it commutes with derivatives. Another 
advantage is that the arising form of the Master BRST identity is equally 
simple as in the model with the auxiliary field $B$. Namely, let 
$F$ be a functional of the fields $A_{\mu},u,\tilde{u}$. The image under $s_0$ 
might depend on $B$, but 
\begin{equation}
  s_0(F)-t(F)=\int dx\,(B(x)-\d A(x))\frac{\delta
    F}{\delta\tilde{u}(x)}
\label{s_0-t}
\end{equation}
and $B-\d A=\frac{\delta S_0}{\delta B}$. Hence if we replace in the 
Master BRST identity $s_0$ by $t$ the correction terms due to the MWI involve 
derivatives with respect to $B$. Hence if none of the functionals 
$F_i$ depends on $B$, we obtain the Master BRST Identity (second form)
\begin{equation}
 \{Q_0,R_{S_0}(F_1,\ldots,F_n\}_{S_0}= \sum_{k=1}^n 
R_{S_0}(F_1,\ldots,t(F_k),\ldots,F_n)\label{mbrst:t}
\end{equation}
where now the field $B$ has been eliminated.

\item Another possibility is to use the fact that on the algebra 
$\mathcal{P}_0=\mathcal{P}/\mathcal{J}$, where $\mathcal{J}$ is the 
ideal of $\mathcal{P}$ which is generated by the free field equations, the 
BRS transformation $\hat{s}_0$ is well defined e.g. by the 
adjoint action of $Q_0$ (w.r.t. the Poisson bracket). Using the section 
$\sigma:\mathcal{P}_0\to\mathcal{P}$ of Sect.~3.2 \footnote{The
  avoidance of the field $B$ contradicts the requirement (vi) on
  $\sigma$, but this is no harm. To do so we choose
  $\sigma \pi (\d_\nu A^\nu)=\d_\nu A^\nu$, cf. (\ref{spi:1}).}
we set
\begin{equation}
\hat{t}\=d\sigma\hat{s_0}\pi =\sigma\pi t, 
\end{equation}
$\pi$ denoting the canonical homomorphism $\mathcal{P}\to\mathcal{P}_0$
(as in Sect.~3.2). $\hat{t}$ has vanishing square but does 
not commute with derivatives. It naturally occurs if one considers the 
entries of retarded (and time-ordered) products as functionals of the 
free fields, as traditionally done in causal perturbation theory.
The price to be paid is a more complicated form of the Master BRST Identity. 
Namely, $(t-\hat{t})(F)=(1-\sigma\pi)t(F)\in\mathcal{J}_{S_0}$, 
hence from the MWI we find the Master BRST identity (third form) 
\begin{gather}
  \{Q_0,R_{S_0}(F_1,\ldots,F_n)\}_{S_0} =
\sum_{k=1}^n 
R_{S_0}(F_1,\ldots,\hat{t}(F_k),\ldots,F_n)
 \notag\\
-\sum_{k\neq l} 
R_{S_0}((F_i)_{i<\max(l,k)},
\delta_{(t-\hat{t})F_k}F_l,(F_j)_{j>\max(l,k)})\ .\label{mbrst:G^1}
\end{gather}
In many applications $\hat t(P),\> P\in\mathcal{P}$, can be written
as the divergence of another field polynomial 
$P^{\prime\nu}\in\mathcal{P}$ by using the free field equations, i.e.
$\pi\hat t(P)\bigl(\equiv \pi t(P)\bigr)=\pi (\d_\nu P^{\prime\nu})$,
or equivalently
\begin{equation}
  \hat t(P)=\sigma\pi (\d_\nu P^{\prime\nu}).
\end{equation}
In the next Subsect. we will see that an admissible interaction
$L_\mathrm{int}$ must fulfil this property. So let us assume that in 
(\ref{mbrst:G^1})
\begin{equation}
  F_k=f_kP_k\quad\mathrm{with}\quad\hat t(P_k)=\sigma\pi 
(\d_\nu P_k^{\prime\nu})\ ,\quad f_k\in\mathcal{D}(\MM),\quad
k=1,...,n.
\end{equation}
In $R_{S_0}(F_1,\ldots,f_k\sigma\pi 
(\d_\nu P_k^{\prime\nu}),\ldots,F_n)$ we would then like to move the
derivative to the test function $f_k$ (i.e. outside of the unsmeared
retarded product). This produces corrections which can directly be read
off from the second formulation of the MWI (\ref{mwi}):
\begin{gather}
  \{Q_0,R_{S_0}(f_1P_1,\ldots,f_nP_n)\}_{S_0} =\notag\\
-\sum_{k=1}^n R_{S_0}(f_1P_1,\ldots,(\d_\nu f_k)\sigma\pi 
(P_k^{\prime\nu}),\ldots,f_nP_n)\notag\\
+\sum_{k\neq l} R_{S_0}((f_iP_i)_{i<\max(l,k)},
G((P_k,P_k')f_k,P_lf_l),(f_jP_j)_{j>\max(l,k)})\ ,\label{mbrst:G}
\end{gather}
where
\begin{equation}
  G((P_1,P_1')f_1,P_2f_2)\=d -\delta_{(t-\hat{t})f_1P_1}(f_2P_2)-
  \sum_{\chi,\psi\in {\cal G}}f_1\>\sigma\pi
        \Bigl(\frac{\d P_{1\nu}'}{\d\chi}\Bigr)\>
        \delta^{\nu}_{\chi,\psi}
        \frac{\d (f_2P_2)}{\d\psi}\ .\label{G}
\end{equation}
Obviously, there 
is also a mixed formula in which the step from (\ref{mbrst:s_0}) (or
(\ref{mbrst:t}), or (\ref{mbrst:G^1})) to 
(\ref{mbrst:G}) is done for some of the factors $F_k$ only (not
for all).
In the forms (\ref{mbrst:G^1}) and
(\ref{mbrst:G}) the Master BRST Identity was found in \cite{DF3} with
$\delta_{(t-\hat{t})F_k}F_l$ corresponding to the terms 
$G^{(1)}(F_k,F_l)$. ($G(\ldots)$ (\ref{G}) denotes the same terms.)
\end{itemize}

Note that the Master BRST Identity
(in either form) is independent of the choice of an interaction 
and is therefore well suited for the formalism of causal perturbation theory
where one aims at finding the retarded (or time ordered) 
products not only for the interaction Lagrangian itself but for 
a whole class of fields. 

Given the free (quantum) gauge fields, requirements on the interaction
are formulated in \cite{DF3}, in particular a suitable form of 
BRS-invariance. These requirements determine
the interaction to a far extent \cite{stora,DS,Scharf}. 
Then it is demonstrated that for
such an interaction the validity of particular cases
of the Master BRST Identity and of ghost
number conservation (which is another consequence of the MWI) suffices
for a construction of the net of local algebras of observables. This
construction yields also a space of physical states and
an explicit formula for the computation of the BRS-transformation of
an arbitrary quantum field.

In that reference the requirements 
expressing BRS-invariance of the interaction
have been motivated by the particular case of 
purely massive gauge models (\ref{[Q,S]=0}) (in which the
adiabatic limit exists, see e.g. \cite{DSchroer,G,DHKS}), and by what
has been used in the construction and holds true in the most important
examples. However, it is desirable to derive these conditions from more
fundamental principles without using the adiabatic limit. 
The MWI and AWI are well suited tools for such a derivation, as it is 
demonstrated in Appendix B. However, in the next Subsection we 
determine the admissible interaction of a {\it local} gauge theory 
independently of the corresponding procedure in \cite{DF3}. This is a
further important application of the MWI and AWI.

\subsection{Admissible interaction}
By an admissible interactions we understand an  
interaction for which a deformed BRS charge $Q$ exists\footnote{For
  simplicity we do not investigate the existence of $Q$ as an {\it
  operator} (which is used for the construction of physical states 
  in \cite{DF} and \cite{DF3}). This existence
  involves an infrared problem which can be avoided by a spatial
  compactification \cite{DF}. Here, we only require that $Q$
  implements a nilpotent (graded) derivation on the interacting fields
  (\ref{s<>hat-s})-(\ref{hat-s}), which is a deformation of the free
  BRS-transformation $s_0$ (\ref{s_0}).}.
Let the free action $S_0$ and the free BRS-current be given.
We make the ansatz
\begin{equation}
   S_{\mathrm{int}}=\sum_{n\ge 1}S_n\lambda^n\label{S_int}
\end{equation}
with $S_n=\int dx g(x)^n\mathcal{L}_n(x)$, $\mathcal{L}_n\in\mathcal{P}$, 
$g\in\mathcal{D}(\MM)$, $g\equiv 1$ on some causally complete 
open region $\mathcal{O}_1$. We want to find a conserved current of the 
interacting theory
\begin{equation}
   j_{\mu}=\sum_{n\ge 0}j_{\mu}^{(n)}\lambda^n\label{BRS-current:int}
\end{equation}
where $j_{\mu}^{(0)}$ is the free BRS current. The
BRS-transformation $s:\mathcal{P}\rightarrow\mathcal{P}$, will
be constructed in the form 
\begin{equation}
s=\sum_{n\ge 0}\int dx\,\tilde s_n(x)\lambda^n\label{s:Reihe}
\end{equation}
with $\tilde s_n(x)$ a local (graded) derivation, and 
$\tilde s_0(x)$ given by
(\ref{s_0(x)}). In addition we require that $s$
is nilpotent on the space of solutions and fulfills
\begin{equation}
   (s(F))_S=\hat{s}(F_S)\ ,\label{s<>hat-s}
\end{equation}
where $\hat{s}$ is defined in terms
of the BRS-current $j$ (\ref{BRS-current:int}) by
\begin{equation}
   \hat{s}(F_S)=\{Q,F_S\} \quad , \ Q\=d\langle j,b\rangle_{S} \quad ,
   \forall \>\mathrm{local}\>F \>\mathrm{with}\>
   \supp F\subset\mathcal{O}\,\label{hat-s}
\end{equation}
(with $S=S_0+S_{\mathrm{int}}$). Thereby,
$\mathcal{O}$ is a causally complete open region with 
$\overline{\mathcal{O}}\subset \mathcal{O}_1$. Due to current
conservation there is a rather large freedom in the choice of 
$b=(b_{\mu})$. It only needs to be a smooth version of a delta function 
on a Cauchy surface of $\mathcal{O}$ with $\supp b\subset
\mathcal{O}_1$. For later purpose we choose $b$ in the following way:
let $h\in\mathcal{D}(\mathcal{O}_1)$ with $h\equiv 1$ on $\mathcal{O}$.
Then the $b_\mu$ which we will use later on is obtained from $\d_\mu
h$ by the same causal splitting as in (\ref{QW}).  
Note that for an arbitrary given local $F\in\mathcal{F(C)}$ the regions 
$\mathcal{O}$, $\mathcal{O}_1$ as well as $h$ and $b$ can be suitably 
adjusted.

We require current conservation within the region $\mathcal{O}_1$
(where $g$ is constant) only,
\begin{equation} 
  R_{S_0}(e_{\otimes}^{S_{\mathrm{int}}},\d j(x))=0 \ ,\ 
  \forall x\in\mathcal{O}_1 \ .\label{curr-conserv}
\end{equation}
To zeroth order in $\lambda$ this is simply the condition that the 
free BRS current is conserved in the free theory
\begin{equation} 
 \d j^{(0)}\equiv :G_0\in\mathcal{J}\label{cc:n=0}
\end{equation}
and apply the MWI,
\begin{equation} 
 0= R_{S_0}(e_{\otimes}^{S_{\mathrm{int}}},\tilde s_0(x)S_{\mathrm{int}} 
  -\sum_{n\ge 1}\d j^{(n)}(x)\lambda^n)\ ,\quad x\in\mathcal{O}_1,
\end{equation}
where $\tilde s_0(x)=\delta_{G_0(x)}$ (\ref{s_0(x)}).

To first order we find the requirement
\begin{equation} 
 \tilde s_0(x)S_1 -\d j^{(1)}(x)\equiv :-G_1(x)\in\mathcal{J}\ .\label{G1}
\end{equation}
Therefore we can apply again the MWI and obtain
\begin{equation} 
 0= R_{S_0}(e_{\otimes}^{S_{\mathrm{int}}},\tilde s_1(x)S_{\mathrm{int}}
 + \tilde s_0(x)\sum_{n\ge 2}S_n\lambda^n
  -\sum_{n\ge 2}\d j^{(n)}(x)\lambda^n)\ ,\quad x\in\mathcal{O}_1,
\end{equation}
with $\tilde s_1(x)=\delta_{G_1(x)}$. Iterating the procedure we obtain the 
conditions
\begin{equation} 
 \tilde s_0(x)S_n+\tilde s_1(x)S_{n-1}+\ldots +\tilde s_{n-1}(x)S_1 -
 \d j^{(n)}(x)\equiv :-G_n(x)\in\mathcal{J}\ ,\label{G_n}
\end{equation}
and set $\tilde s_n(x):=\delta_{G_n(x)}$. We see that at every order, 
$S_n$ must be chosen such that
\begin{equation} 
  \sum_{k=0}^{n-1} \tilde s_k(x)S_{n-k}\in 
\mathcal{P}_{\mathrm{div}}+\mathcal{J}
\end{equation}
where $\mathcal{P}_{\mathrm{div}}=\{\d^{\mu}f_{\mu}\>,\>
f_{\mu}\in\mathcal{P}\}$. This inductive determination of $s$ and 
$S_\mathrm{int}$ by requiring $\d j=0$ has some similarity with the
procedure in \cite{Hurth}, cf. Appendix B. 
Since $G_n(x)=\delta_{G_n(x)}S_0=\tilde s_n(x)S_0$
the relation (\ref{G_n}) and current conservation imply 'local' 
BRS-invariance of the action 
$S=S_0+S_{\mathrm{int}}$ within $\mathcal{O}_1$:
\begin{equation}
  \tilde s(x)S=\sum_{n\ge 0}\lambda^n\sum_{k=0}^n\tilde s_k(x)S_{n-k}=
\d j(x)\ ,\quad x\in\mathcal{O}_1\ ,\label{sS}
\end{equation}
and hence
\begin{equation}
  \bigl(\tilde s(x)S\bigr)_S=0\ ,\quad\quad\forall x\in\mathcal{O}_1\ .
\end{equation}

It remains to verify that the so constructed BRS-transformation $s$
(\ref{s:Reihe}) satisfies (\ref{s<>hat-s}) and the nilpotency
\begin{equation} 
 0=(s^2(F))_S=\hat{s}^2(F_S) \quad \forall F \ .
\end{equation}
To prove the first property we choose $h$ and $b$ as in
(\ref{QW}). Analogously to (\ref{QW1}) we find
\begin{equation} 
 \{Q,F_S\}=-R_S(\langle \d j, h\rangle,F )\ .
\end{equation}
Because $\langle \d j, h\rangle\in \mathcal{J}_S$ we can apply the MWI:
\begin{equation} 
 \{Q,F_S\}=(\delta_{\langle \d j, h\rangle} F)_S=(s(F))_S\ .
\end{equation}
In the last step we have used (\ref{sS})
as well as $\supp h\subset\mathcal{O}_1$ and $h\equiv 1$ on $\supp F$.

Finally, we want to check the nilpotency of $\hat{s}$. 
Using the Jacobi identity we find
\begin{gather} 
 (s^2(F))_S=\frac{1}{2}\{Q(b),\{Q(b'),F_S\}\} 
           +\frac{1}{2}\{Q(b'),\{Q(b),F_S\}\} \notag\\
           = \{\{Q(b'),Q(b)\},F_S\} 
\end{gather}
for all admissible test functions $b,b'$ (depending on the support of $F$).
We may now choose  $b,b'$ such that $b'$ satisfies the conditions also 
with respect to the support of $b$ (and of course $\supp b'
\subset\mathcal{O}_1$). Then
\begin{equation} 
 (s^2(F))_S=\{s(\langle j,b\rangle)_S,F_S\}\ . 
\end{equation}
We now assume that 
\begin{equation} 
  s(j_{\mu})=\d^{\nu}C_{\mu\nu}+H_{\mu}
  \label{nilpotent}
\end{equation}
with an antisymmetric tensor field
$C_{\mu\nu}\in\mathcal{P}$ and $H_{\mu}\in\mathcal{J_S}$.
Then
\begin{equation} 
 (s^2(F))_S=\{\langle\d^{\nu}C_{\mu\nu},b^{\mu}\rangle_S,F_S\}
           =\frac{1}{2}\{\langle C_{\mu\nu},
            \d^{\mu}b^{\nu}-\d^{\nu}b^{\mu}\rangle_S,F_S\}=0
\end{equation}
since the support of $(\d^{\nu}b^{\mu}-\d^{\mu}b^{\nu})$ is spacelike
to the support of $F$.\footnote{In $\supp F+\bar{V}_\pm$ we have $b=0$ or
$b^\mu=\d^\mu h$.}

For massless gauge fields without matter fields (i.e. the model
studied in the preceding Subsect.)
the usual expression for the BRS-current
\begin{equation} 
  j^\mu = B\cdot D^\mu u-\d^\mu B\cdot u +\frac{1}{2}\d^\mu\tilde u
\cdot (u\times u)
\end{equation}
(where $(D^\mu u)_a= \d^\mu u_a + f_{abc}A^\mu_b u_c$) is
BRS-invariant: $s(j^\mu)=0$. So the assumption (\ref{nilpotent})
is trivially satisfied. 

In cases where the condition (\ref{nilpotent}) cannot be 
directly checked one may use a perturbative formulation. Set
\begin{equation} 
  H\=d s(j)-\d C
\end{equation}
for some choice of $C=(C_{\mu\nu})$. The condition $H\in\mathcal{J_S}$
means that
\begin{equation} 
  R_{S_0}(e_{\otimes}^{S_{\mathrm{int}}},H)=0  \ .
\end{equation}
for all $\lambda$. In zeroth order we find that $H^{(0)}\in\mathcal{J}_{S_0}$.
Set $K^{(0)}=-H^{(0)}$. We apply the MWI and get
\begin{equation} 
  R_{S_0}(e_{\otimes}^{S_{\mathrm{int}}},
  \delta_{K^{(0)}}S_{\mathrm{int}}+H^{(n\ge 1)})=0  \ .
\end{equation}
In lowest order this implies
\begin{equation} 
  K^{(1)}\=d -\delta_{K^{(0)}}S_{1}-H^{(1)}\in\mathcal{J}_{S_0} \ .
\end{equation}
We now define recursively
\begin{equation} 
  K^{(n)}\=d -\sum_{k=1}^n\delta_{K^{(n-k)}}S_{k}-H^{(n)}
\end{equation}
and prove by induction that
\begin{equation} 
  R_{S_0}(e_{\otimes}^{S_{\mathrm{int}}},
  \sum_{k=1}^n\delta_{K^{(n-k)}}S_{l\ge k}+H^{(l\ge n)})=0  \ .
\label{iteration}
\end{equation}
The lowest order term of (\ref{iteration}) is $(-K^{(n)})_{S_0}$, hence 
$K^{(n)}\in\mathcal{J}_{S_0}$.
The recursion problem can be solved if for every $n$ there exists an 
antisymmetric tensor field $C^{(n)}$ and a vector field 
$K^{(n)}_{\mu}\in\mathcal{J}_{S_0}$ such that
\begin{equation} 
  \sum_{k=1}^n\delta_{K^{(n-k)}}S_{k}+
\sum_{k=0}^n s_{n-k}(j^{(k)})=\d C^{(n)}-K^{(n)}\ .
\end{equation}
\bigskip

In the given derivation we have used various cases of the MWI.
In QFT it may therefore happen that the appearance of anomalies
restricts the set of admissible interactions further, e.g.
models with (non-compensated) axial anomalies must be excluded.

The conditions on a gauge interaction found here differ somewhat from
the corresponding conditions in \cite{DF3} (cf. Appendix B) or in
\cite{Scharf}. For example: to ensure renormalizability it is required
$S_l=0$ for all $l\geq 3$ in \cite{DF3}. And the condition 
(\ref{curr-conserv:M}) (which is the input for the derivation of the
conditions of \cite{DF3} given in Appendix B) is stronger than
(\ref{curr-conserv}). However, for the class of renormalizable
(by power counting) interactions we expect that the 
requirements derived here and the ones given in \cite{DF3}
have precisely the same solutions.
\section{Appendix A: Construction of the map $\sigma$}
We work in $d=4$ dimensions. In the first part we construct 
recursively a particular $\sigma$ in the enlarged model (i.e.
with the field $\varphi^\mu$). This construction applies also to
the non-enlarged model. For the latter we prove that $\sigma$
is unique (in the second part of this Appendix).
\subsection{Particular solution for $\sigma$}
We define
\begin{equation}
  H_{s,n}^{\mu_1...\mu_s}\=d\square^n\d^{\mu_1}...\d^{\mu_s}\varphi-
(-m^2)^n\sigma\pi(\d^{\mu_1}...\d^{\mu_s}\varphi),\label{H}
\end{equation}
which is obviously an element of the ideal $\mathcal{J}$ 
(\ref{eq:ideal of free fields}) and 
totally symmetrical in $\mu_1,...,\mu_s$. Hence, these properties hold 
true also for
\begin{gather}
  F_{r,n}^{\nu_1...\nu_r}=\frac{(-1)^n}{N_{r,n}}\sum_{1\leq j_1<...
<j_{2n}\leq  r}\sum_{\pi\in\mathcal{S}_{2n}}\frac{1}{2^n n!}
g^{\nu_{j_{\pi 1}}\nu_{j_{\pi 2}}}...\notag\\
...g^{\nu_{j_{\pi (2n-1)}}\nu_{j_{\pi 2n}}}
H_{r-2n,n}^{\nu_1...\hat j_1...\hat j_{2n}...\nu_r},\quad n=1,...,
\Bigl[\frac{r}{2}\Bigr],\label{F_r,n}
\end{gather}
(the hat means that the corresponding index is omitted,
and $[r/2]=\frac{r}{2}$ if $r$ even
and $[r/2]=\frac{r-1}{2}$ if $r$ odd) where
\begin{equation}
  N_{r,n}=\prod_{l=1}^n(2-2l+2r).\label{N}
\end{equation}
We now claim that
\begin{equation}
\sigma\pi (\d^{\nu_1}...\d^{\nu_r}\varphi)
=\sigma\pi(\d^{\nu_1}...\d^{\nu_{r-1}}\varphi^{\nu_r})=
\d^{\nu_1}...\d^{\nu_r}\varphi+\sum_{n=1}^{[\frac{r}{2}]}
F_{r,n}^{\nu_1...\nu_r}\label{spi}
\end{equation}
yields a particular solution for $\sigma\pi (\d^{\nu_1}...
\d^{\nu_r}\varphi)$ for $r\geq 2$ by recursion with respect to $r$.
Together with the formulas (\ref{spi:0})-(\ref{spi:1}) for $r=0,1$
this determines a map $\sigma$ completely.\\
\textit{Proof}: The only non-trivial point is to verify $\sigma\pi
(\d^a\d_\mu\varphi^\mu)=-m^2\sigma\pi (\d^a\varphi)$. 
Because the r.h.s. of (\ref{spi})
is totally symmetrical in $\nu_1,...,\nu_r$, it suffices to show
\begin{equation}
  g_{\nu_1\nu_2}\sigma\pi (\d^{\nu_1}...\d^{\nu_r}\varphi)=
-m^2 \sigma\pi (\d^{\nu_3}...\d^{\nu_r}\varphi).\label{spi:eq}
\end{equation}
For this purpose we set $N_{r,0}\=d 1$ and
\begin{gather}
  G_{r,s}^{\nu_3...\nu_r}\=d\sum_{3\leq j_1<...
<j_{2s}\leq  r}\sum_{\pi\in\mathcal{S}_{2s}}\frac{1}{2^s s!}
g^{\nu_{j_{\pi 1}}\nu_{j_{\pi 2}}}...\notag\\
...g^{\nu_{j_{\pi (2s-1)}}\nu_{j_{\pi 2s}}}
H_{r-2(s+1),s+1}^{\nu_3...\hat j_1...\hat j_{2s}...\nu_r}\label{G_rs}
\end{gather}
for $0\leq 2s\leq r-2$, and $G_{r,s}\=d 0$ for $2s> r-2$. By
inserting the definitions we find that the identity
\begin{equation}
  g_{\nu_1\nu_2}F_{r,n}^{\nu_1...\nu_r}=\frac{(-1)^n}{N_{r,n-1}}
G_{r,n-1}^{\nu_3...\nu_r}+\frac{(-1)^n}{N_{r,n}}
G_{r,n}^{\nu_3...\nu_r},\quad n=1,...,
\Bigl[\frac{r}{2}\Bigr],\label{F:kontr}
\end{equation}
implies the assertion (\ref{spi:eq}).

To prove (\ref{F:kontr}) we write $F_{r,n}^{\nu_1...\nu_r}$ in the
following form:
\begin{gather}
  F_{r,n}^{\nu_1...\nu_r}=\frac{(-1)^n}{N_{r,n-1}(2-2n+2r)}\Bigl[
g^{\nu_1\nu_2}G_{r,n-1}^{\nu_3...\nu_r}+\notag\\
\sum_{3\leq l<k\leq r}(g^{\nu_1\nu_l}g^{\nu_2\nu_k}+
g^{\nu_1\nu_k}g^{\nu_2\nu_l})\sum g^{...}...H_{r-2n,n}+\notag\\
\sum_{3\leq l\leq r}[g^{\nu_1\nu_l}\sum g^{...}...
H_{r-2n,n}^{\nu_2...}+(\nu_1\leftrightarrow\nu_2)]\Bigr]\notag\\
+\frac{(-1)^n}{N_{r,n}}\sum g^{...}...H_{r-2n,n}^{\nu_1\nu_2...}.
\end{gather}
Contraction of the second (third respectively) line with 
$g_{\nu_1\nu_2}$ gives $2(n-1)G_{r,n-1}^{\nu_3...\nu_r}$
(and $2(r-2n)G_{r,n-1}^{\nu_3...\nu_r}$ resp.). In the last line we
use
\begin{equation}
  g_{\mu_1\mu_2}H_{s,n}^{\mu_1...\mu_s}=H_{s-2,n+1}^{\mu_3...\mu_s},
\end{equation}
and end up with
\begin{gather}
g_{\nu_1\nu_2}F_{r,n}^{\nu_1...\nu_r}=\frac{(-1)^n}{N_{r,n-1}(2-2n+2r)}
(4+2(n-1)+2(r-2n))G_{r,n-1}^{\nu_3...\nu_r}\notag\\
+\frac{(-1)^n}{N_{r,n}}G_{r,n}^{\nu_3...\nu_r}.\quad\quad\quad\w
\end{gather}
\subsection{Uniqueness of $\sigma$ for a single 
real Klein-Gordon field}
We will prove that the map $\sigma$ is unique for $\mathcal{P}$ and
$\mathcal{J}$ given by (\ref{P}) and (\ref{eq:ideal of free fields}).
Obviously, the relations $\sigma\pi 
(\varphi)=\varphi$, $\sigma\pi (\d^\mu\varphi)=\d^\mu\varphi$ and the 
recursive formula (\ref{spi}) give a solution for $\sigma$ which we
denote by $\sigma_0$. Analogously to (\ref{spi:ansatz}) and by the
requirements (iii) and (v) on $\sigma$ we conclude that the most 
general solution for $\sigma$ is of the form:
\begin{gather}
  \sigma\pi (\d^{\nu_1}...\d^{\nu_r}\varphi)=
\sigma_0\pi (\d^{\nu_1}...\d^{\nu_r}\varphi)+M^{\nu_1...\nu_r},
\notag\\
M^{\nu_1...\nu_r}\=d\frac{1}{r!}\sum_{\pi\in\mathcal{S}_r}
\sum_{l=1}^{[r/2]}\sum_{j=1}^l a_{l,j}g^{\nu_{\pi 1}\nu_{\pi2}}
...g^{\nu_{\pi (2l-1)}\nu_{\pi 2l}}\d^{\nu_{\pi (2l+1)}}...
\d^{\nu_{\pi r}}\square^{(l-j)}(\square+m^2)\varphi,
\end{gather}
where $a_{l,j}\in\RR$ is a constant. An $\epsilon$-tensor is excluded
in $M^{\nu_1...\nu_r}$ by the total symmetry in $\nu_1,...,\nu_r$.
We are now going to show that $g_{\nu_{r-1}\nu_r}M^{\nu_1...\nu_r}=0$
(see (\ref{spi:eq})) yields $a_{l,j}=0$, $\forall l,j$. We use the
equation
\begin{gather}
 g_{\nu_{r-1}\nu_r} \frac{1}{r!}\sum_{\pi\in\mathcal{S}_r}
g^{\nu_{\pi 1}\nu_{\pi2}}...g^{\nu_{\pi (2l-1)}\nu_{\pi 2l}}
\d^{\nu_{\pi (2l+1)}}...\d^{\nu_{\pi r}}=\notag\\
\frac{1}{(r-2)!}\sum_{\pi\in\mathcal{S}_{r-2}}\{
N_{r,l}g^{\nu_{\pi 1}\nu_{\pi2}}...g^{\nu_{\pi (2l-3)}\nu_{\pi (2l-2)}}
\d^{\nu_{\pi (2l-1)}}...\d^{\nu_{\pi (r-2)}}\notag\\
+M_{r,l}g^{\nu_{\pi 1}\nu_{\pi2}}...g^{\nu_{\pi (2l-1)}\nu_{\pi 2l}}
\d^{\nu_{\pi (2l+1)}}...\d^{\nu_{\pi (r-2)}}\square\}
\end{gather}
where $N_{r,l}$ and $ M_{r,l}$ are non-vanishing combinatorical
factors, except $M_{r,[r/2]}=0$. The requirement $g_{\nu_{r-1}\nu_r}
M^{\nu_1...\nu_r}=0$ gives the following chains of equations
\begin{equation}
  a_{s,s}N_{r,s}=0,\quad\quad a_{l,s}M_{r,l}+a_{l+1,s}N_{r,l+1}=0
\quad\forall l\in\{s,s+1,...,[r/2]-1\},
\end{equation}
where the chains are labeled by $s=1,2,...,[r/2]$. We find indeed 
$a_{l,j}=0$, $\forall l,j$. $\quad\w$
\section{Appendix B: Formulation of BRS-invariance of the interaction
used in \cite{DF3}}
To derive the conditions which are used in 
\cite{DF3} to express BRS-invariance of the interaction\footnote{We
shall not obtain that conditions in full generality. However, as
far as we know, the particular version which we shall obtain
determines the interaction to the same extent, see below.} 
we only require current conservation (\ref{curr-conserv}) but
generalized to all $x\in\MM$. The latter makes possible an integration
over $x\in\MM$, which removes $\d j^{(n)}$ from (\ref{G_n}) and replaces
$\tilde s_0(x)$ by $s_0$. This generalization is done by admitting
that current conservation at $x$ is violated by a term of the form
$M_\nu (x)\d^\nu g(x)$. Since we want to obtain conditions on the
interaction which are expressable in terms of {\it free} fields, we
require that $M_\nu$ is in the range of $\sigma$: $M_\nu =\sigma
\pi (K_\nu)$ for some $K_\nu$. In detail our input is the requirement
that there exist an
interaction $S_\mathrm{int}$ (\ref{S_int}), a BRS-current $j_\mu$
(\ref{BRS-current:int}) (with compact support\footnote{Usually 
we expect $\supp j_\mu^{(n)}\subset\supp g$ for $n\geq 1$.}
of $j_\mu^{(n)}\>\forall n\geq 1$) and a formal power series
\begin{equation} 
  K_\nu (x) =\sum_{n\geq 1}\lambda^n K^{(n)}_\nu (x),
\quad\quad K^{(n)}_\nu\in\mathcal{P}\otimes\mathcal{C}(\MM)\ ,
\end{equation}
such that
\begin{equation} 
R_{S_0}(e_{\otimes}^{S_{\mathrm{int}}},\d j(x)-\sigma\pi (K_\nu)(x)
\d^\nu g(x))=0 \ ,\ 
  \forall x\in\MM \ ,\ \forall g\in\mathcal{D}(\MM)\ .
\label{curr-conserv:M}
\end{equation}
So, in a formalism with auxiliary fields, $S_\mathrm{int}$ and $j$
depend on the choice of $\sigma$, but this dependence drops out 
when the auxiliary fields are eliminated.\footnote{We explain this
  for the 4-gluon interaction in the formalism of \cite{DF3}. For a
  $\sigma$ corresponding to  $C_A=-\frac{1}{2}$ this coupling is
  generated by $S_1$, but for $C_A=0$ it appears in $S_2$, cf. the
  Remark at the end of Sect.~4.} The condition 
(\ref{curr-conserv:M}) corresponds to the 'Quantum Noether
condition in terms of interacting fields' \cite{Hurth}, however  
the formalisms are quite different.

In working out the consequences of (\ref{curr-conserv:M}) we will
frequently use the AWI and the MWI (in particular various formulations
of the Master BRST Identity) without mentioning it. As in \cite{DF3}
we use a formalism without $B$-field; so we may replace $s_0$ by $t$
due to (\ref{s_0-t}). Similarly to (\ref{G_n}) our requirement
(\ref{curr-conserv:M}) is equivalent to the sequence of conditions
\begin{gather} 
 \tilde s_0(x)S_n+\tilde s_1(x)S_{n-1}+\ldots +\tilde s_{n-1}(x)S_1 -
 \d j^{(n)}(x)+\sigma\pi (K^{(n)}_\nu)(x)
\d^\nu g(x)\notag\\
\equiv :-G_n(x)\in\mathcal{J}\ ,\ n\geq 1\ ,\label{G_n:M}
\end{gather}
and (\ref{cc:n=0}) for $n=0$, where $\tilde s_n(x)\=d\delta_{G_n(x)}$.
We will proceed in the following way: first we insert formulas which 
are inductively known into 
$\tilde s_1(x)S_{n-1}+\ldots +\tilde s_{n-1}(x)S_1 $. Then we
integrate the resulting equation over $x\in\MM$. Thereby, we take into
account that $j^{(n)}$ has compact support.

\begin{itemize}
\item To first order we obtain
\begin{equation}
    (tS_1)_{S_0}=(\d^\nu\,\sigma\pi(K^{(1)}_\nu))(g)_{S_0}=
    \sigma\pi(\d^\nu K^{(1)}_\nu)(g)_{S_0}\ .
\end{equation}
Since $g$ is arbitrary we end up with the condition that there must
exist $\mathcal{L}_1,\, K^{(1)}_\nu\in\mathcal{P}$ with
\begin{equation}
  \hat t \mathcal{L}_1=\sigma\pi(\d^\nu K^{(1)}_\nu)\ .\label{cond:1}
\end{equation}

\item To second order $n=2$ we multiply equation (\ref{G_n:M}) by 2.
Then we insert once 
\begin{gather}
  (\tilde s_1(x)S_1)_{S_0}=-R_{S_0}(G_1(x),S_1)\notag\\
=-R_{S_0}(\d j^{(1)}(x),S_1)
+R_{S_0}(\tilde s_0(x)S_1,S_1)+R_{S_0}(\sigma\pi(K^{(1)}_\nu)(x)\d^\nu
g(x),S_1)\label{s_1S_1}
\end{gather}
and once $(\tilde
s_1(x)S_1)_{S_0}=-R_{S_0}(S_1,G_1(x))=...\ .$ Integration yields
\begin{equation}
  \{Q_0,R^N_{S_0}(S_1,S_1)\}=-R^N_{S_0}(\sigma\pi(K^{(1)}_\nu)\d^\nu
g,S_1)-R^N_{S_0}(S_1,\sigma\pi(K^{(1)}_\nu)\d^\nu g)\ ,\label{Q,R(S,S)}
\end{equation}
where
\begin{equation}
  R^N_{S_0}(S_1,S_1)\=d R_{S_0}(S_1,S_1)+2(S_2)_{S_0}\label{R^N(S,S)}
\end{equation}
and 
\begin{gather}
  (R^N_{S_0}-R_{S_0})(\sigma\pi(K^{(1)}_\nu)f,S_1):\=d 
\sigma\pi(K^{(2)}_\nu)(f)_{S_0}\notag\\
\=d :(R^N_{S_0}-R_{S_0})
(S_1,\sigma\pi(K^{(1)}_\nu)f)\ .\label{R^N(K,S)}
\end{gather}
are shorthand notations which we will interpret below.
We find the additional requirement that there must exist
$\mathcal{L}_2\in\mathcal{P},\> K^{(2)}_\nu
\in\mathcal{P}\otimes\mathcal{C}(\MM)$ with
\begin{equation}
   \hat t \mathcal{L}_2(g^2)+\sigma\pi G((\mathcal{L}_1,K^{(1)})g,
\mathcal{L}_1g)+\sigma\pi(K^{(2)}_\nu)(\d^\nu g)=0\ .\label{cond:2}
\end{equation}

\item To third order $n=3$ we multiply equation (\ref{G_n:M}) by
  $3!=6$. Then we insert twice $(\tilde s_1(x)S_2)_{S_0}=-
R_{S_0}(S_2,G_1(x))=...$ (cf. (\ref{s_1S_1})) and
\begin{gather}
  (\tilde s_2(x)S_1)_{S_0}=-R_{S_0}(G_2(x),S_1)
=-R_{S_0}(\d j^{(2)}(x),S_1)\notag\\
+R_{S_0}(\tilde s_0(x)S_2,S_1)+R_{S_0}(\delta_{G_1(x)}S_1,S_1)
+R_{S_0}(\sigma\pi(K^{(2)}_\nu)(x)\d^\nu g(x),S_1)\ ,\label{s_2S_1}
\end{gather}
as well as four times $(\tilde s_1(x)S_2)_{S_0}=-
R_{S_0}(G_1(x),S_2)=...$ and $(\tilde s_2(x)S_1)_{S_0}=-R_{S_0}
(S_1,G_2(x))=...\ .$ Next we use 
\begin{gather}
  R_{S_0}(\delta_{G_1(x)}S_1,S_1)+R_{S_0}(S_1,\delta_{G_1(x)}S_1)=
-R_{S_0}(G_1(x),S_1,S_1)\notag\\
=-R_{S_0}(\d j^{(1)}(x),S_1,S_1)+R_{S_0}(\tilde s_0(x)S_1,S_1,S_1)\notag\\
+R_{S_0}(\sigma\pi(K^{(1)}_\nu)(x)\d^\nu g(x),S_1,S_1)
\end{gather}
and $R_{S_0}(S_1,\delta_{G_1(x)}S_1)=-\frac{1}{2}R_{S_0}
(S_1,S_1,G_1(x))=...\ .$ By integration we obtain
\begin{gather}
  \{Q_0,R^N_{S_0}(S_1,S_1,S_1)\}=\notag\\
-2R^N_{S_0}(\sigma\pi(K^{(1)}_\nu)\d^\nu
g,S_1,S_1)-R^N_{S_0}(S_1,S_1,\sigma\pi(K^{(1)}_\nu)\d^\nu g)\ ,
\label{Q,R(S,S,S)}
\end{gather}
where
\begin{equation}
  R^N_{S_0}(S_1,S_1,S_1)\=d R_{S_0}(S_1,S_1,S_1)+2R_{S_0}(S_2,S_1)
+4R_{S_0}(S_1,S_2)+6(S_3)_{S_0}\label{R^N(S,S,S)}
\end{equation}
and 
\begin{gather}
  (R^N_{S_0}-R_{S_0})(\sigma\pi(K^{(1)}_\nu)f,S_1,S_1):\=d 
2R_{S_0}(\sigma\pi(K^{(1)}_\nu)f,S_2)\notag\\
+R_{S_0}(\sigma\pi(K^{(2)}_\nu)f,
S_1)+R_{S_0}(S_1,\sigma\pi(K^{(2)}_\nu)f)+2\sigma\pi(K^{(3)}_\nu)
(f)\ ,\notag\\
(R^N_{S_0}-R_{S_0})(S_1,S_1,\sigma\pi(K^{(1)}_\nu)f):\=d \notag\\
2R_{S_0}(S_2,\sigma\pi(K^{(1)}_\nu)f)+2R_{S_0}(S_1,\sigma\pi(K^{(2)}_\nu)f)
+2\sigma\pi(K^{(3)}_\nu)(f)\ .\label{R^N(S,S,K)}
\end{gather}
With $H\=d tS_2+G((\mathcal{L}_1,K^{(1)})g,
\mathcal{L}_1g)+\sigma\pi(K^{(2)}_\nu)(\d^\nu g)$ 
and by using also a mixed version of the Master BRST Identity,
\begin{gather}
  \{Q_0,R_{S_0}(S_1,S_2)\}=-R_{S_0}(\sigma\pi(K^{(1)}_\nu)\d^\nu
g,S_2)\notag\\
+R_{S_0}(S_1,tS_2)+G((\mathcal{L}_1,K^{(1)})g,S_2)_{S_0}\ ,
\end{gather}
we find the following additional requirement: there must exist
$\mathcal{L}_3,\, K^{(3)}$ with
\begin{equation}
   (t \mathcal{L}_3(g^3))_{S_0}+G((\mathcal{L}_1,K^{(1)})g,
\mathcal{L}_2g^2)_{S_0}+\sigma\pi(K^{(3)}_\nu)(\d^\nu g)_{S_0}
-(\delta_HS_1)_{S_0}=0\ ,\label{cond:3a}
\end{equation}
where we have used $H\in\mathcal{J}_{S_0}$ (\ref{cond:2}). For the
models studied in \cite{DF3} it holds $G((\mathcal{L}_1,K^{(1)})g,
\mathcal{L}_2g)\in {\rm ran\>} \sigma$ and it is possible to fulfill
(\ref{cond:3a}) with $\mathcal{L}_3=0= K^{(3)}$. With that we may write
$H=(t-\hat t)S_2$ by using (\ref{cond:2}), and the condition 
(\ref{cond:3a}) reduces to
\begin{equation}
  -\pi\,\delta_{(t-\hat t)S_2}S_1+\pi\,G((\mathcal{L}_1,K^{(1)})g,
\mathcal{L}_2g^2)=0\ .\label{cond:3}
\end{equation}

\item In general it may happen that the higher orders $n\geq 4$ of
(\ref{curr-conserv:M}) restrict $\mathcal{L}_1,\>\mathcal{L}_2$
and $\mathcal{L}_3$ further and non-vanishing expressions for 
$\mathcal{L}_j$ and $K^{(j)}$ are possible for arbitrary high $j$.
We do not work this out here.
Because for the models treated in \cite{DF3} any solution
$(\mathcal{L}_1,\>\mathcal{L}_2,\>\mathcal{L}_3=0,\> K^{(1)},\>
 K^{(2)},\>  K^{(3)}=0)$ of (\ref{cond:1}), (\ref{cond:2}) and 
(\ref{cond:3}) fulfills (\ref{curr-conserv:M}) to all higher orders
$n\geq 4$ with $\mathcal{L}_l=0= K^{(l)},\> \forall l\geq 4$ (up to anomalies,
i.e. violations of the MWI and the AWI).
\end{itemize}

In \cite{DF3} {\it generalizations} of (\ref{cond:1}), (\ref{cond:2}) and 
(\ref{cond:3}) have been used to determine the interaction, namely
(141)\footnote{More precisely we mean here (141) and the antisymmetry
of the ${\cal L}_2^{\mu\nu}$ which appears in that formula.},
(209) and (216) in the published version.
However, at least for the models studied in \cite{DF3} it seems 
that any solution of (\ref{cond:1}), (\ref{cond:2}) and (\ref{cond:3})
with $\mathcal{L}_3=0= K^{(3)}$ satisfies also these generalizations.

So far this Appendix applies to classical field theory and QFT. The 
following discussions are restricted to QFT. The definition
$R^N_{S_0}(h)\=d R_{S_0}(h)$ for $h=S_1,\sigma\pi (K^{(1)}_\nu)f$
(zeroth order), and the above given first order
(\ref{R^N(S,S)})-(\ref{R^N(K,S)}) and second order definitions
(\ref{R^N(S,S,S)})-(\ref{R^N(S,S,K)}) of $R^N_{S_0}$ are compatible
with the main properties of a retarded product. By the latter we mean 
linearity, symmetry of $R_{n,1}$ in the first $n$ entries, causal 
support (\ref{supp(R)}) and in particular the recursion given by 
the GLZ-relation. The continuation (by analogy) of our
inductive evaluation of (\ref{curr-conserv:M}) yields the 
definitons of $R^N_{S_0}(S_1^{\otimes n},S_1)$, $R^N_{S_0}
(S_1^{\otimes n},\sigma\pi (K^{(1)}_\nu)f)$ and
$R^N_{S_0}(\sigma\pi (K^{(1)}_\nu)f\otimes S_1^{\otimes n-1},S_1)$
for all $n\geq 3$. We expect that these whole sequences are compatible
with the (just mentioned) main properties of a retarded product. 
If this holds true we can proceed similarly to formula (214) 
in \cite{DF3} (which is a particular simple case of Theorem 3.1
in \cite{Pinter}): to all orders
we extend $R^N_{S_0}$ to arbitrary factors such 
that $R^N_{S_0}$ agrees with $R_{S_0}$ as far as possible and that
$R_{S_0}\longrightarrow R^N_{S_0}$ is an {\it admissible 
finite renormalization} (i.e. the main properties of a retarded 
product are preserved in this replacement). However, in
general the $R^N_{S_0}$ violate some normalization conditions,
in particular (\ref{eq:N3class}) (i.e. (N3)), the AWI and the MWI,
see \cite{DF3}. With $R^N_{S_0}$ being an admissible 
retarded product, the equations
(\ref{Q,R(S,S)}) and (\ref{Q,R(S,S,S)}) are {\it perturbative gauge
invariance} (in the sense of \cite{DHKS,DS,Scharf}) to second and 
third order. So, 
we have shown to lowest orders that our condition
(\ref{curr-conserv:M}) implies that the interaction is such that
perturbative gauge invariance can be fulfilled by admissible finite
renormalizations. From the requirement that the latter property
holds true one can
derive the interaction (including the Lie-algebraic structure
\cite{stora}) of massless and massive spin-1 gauge fields and the
couplings of spin-2 gauge theories (for an overview see \cite{Scharf}).

To interpret the $R^N_{S_0}$-products in terms of simple equations we
go over to the corresponding time-ordered 
products\footnote{We are not aware of a good notion of 'time-ordered
products' in classical field theory. But in QFT the retarded products
$\{R_{n,1}\> |\> 0\leq n\leq N\}$ determine uniquely the corresponding 
time-ordered products $\{T_n\> |\> 1\leq n\leq N+1\}$ and vice versa,
see e.g. \cite{EG}.} $T^N$. As far as they are determined by 
(\ref{R^N(S,S)})-(\ref{R^N(K,S)}) and 
(\ref{R^N(S,S,S)})-(\ref{R^N(S,S,K)}) they fulfil
\begin{gather}
  T(e_{\otimes}^{S_{\mathrm{int}}})=T^N(e_{\otimes}^{S_1})\ ,\notag\\
T(e_{\otimes}^{S_{\mathrm{int}}}\otimes \sum_{n\geq 1}\lambda^n 
\sigma\pi (K^{(n)}_\nu)f)=T^N(e_{\otimes}^{S_1}\otimes\sigma\pi 
(K^{(1)}_\nu)f)\label{T^N}
\end{gather}
as formal power series in $\lambda$, and we expect that this holds true
also for the higher orders. But understood as series with
respect to the order $n$ of $T_n$ and $T_n^N$, the equations 
(\ref{T^N}) express a reordering: the contributions to $T_n$
of the terms $S_l,\>\sigma\pi (K_\nu^{(l)})$ with $l\geq 2$,
appear on the r.h.s. in time ordered products $T^N_m$ of higher
orders: $(m-n)\geq (l-1)$. The $R^N_{S_0}
(S_1^{\otimes n},\sigma\pi (K^{(1)}_\nu)f)$ satisfy
\begin{equation}
  R_{S_0}(e_{\otimes}^{S_{\mathrm{int}}},\sum_{n\geq 1}
\lambda^n\sigma\pi (K^{(n)}_\nu)f)=R^N_{S_0}
(e_{\otimes}^{S_1},\sigma\pi (K^{(1)}_\nu)f)\ ,\label{R^N}
\end{equation}
however the corresponding relations for $R^N_{S_0}
(S_1^{\otimes n},S_1)$ and $R^N_{S_0}
(\sigma\pi (K^{(1)}_\nu)f\otimes S_1^{\otimes n-1},S_1)$ are
more involved.

\vskip0.5cm
{\bf Acknowledgements:} We profitted from several, very interesting
and detailed letters from Raymond Stora. We also thank Karl-Henning
Rehren for valuable comments, and Tobias Hurth and Kostas Skenderis
for discussions.

\end{document}